\documentclass[lettersize,journal]{IEEEtran}
\usepackage{amsmath,amsfonts}
\usepackage{array}
\pdfoutput=1
\usepackage[caption=false,font=normalsize,labelfont=sf,textfont=sf]{subfig}
\usepackage{textcomp}
\usepackage{stfloats}
\usepackage{url}
\usepackage{verbatim}
\usepackage{graphicx}
\usepackage{diagbox}
\usepackage{algorithm}

\usepackage{algpseudocode}
\usepackage{xcolor}
\usepackage{verbatim}
\usepackage{adjustbox}  
\usepackage{xcolor}  
\usepackage{booktabs}  
\usepackage{makecell}
\usepackage{amssymb}
\usepackage{subcaption}

\def\BibTeX{{\rm B\kern-.05em{\sc i\kern-.025em b}\kern-.08emT\kern-.1667em\lower.7ex\hbox{E}\kern-.125emX}}
\usepackage{balance}
\begin{document}
	\title{
	ISAC Empowered Air-Sea Collaborative System: A UAV-USV Joint Inspection Framework
	}
	\author{
		Rui Zhang, Fuwang Dong, \textit{Member, IEEE,} Wei Wang, \textit{Senior Member, IEEE}
		\thanks{ 
			
			This work was supported by the National Natural Science Foundation of China
			under Grant 62271163. (\textit{Corresponding author: Fuwang Dong.})
			
			The authors are with the College of Intelligent Systems Science and Engineering, Harbin Engineering University, Harbin 150001,
			China (e-mail: azhangrui407@hrbeu.edu.cn; dongfuwang@hrbeu.edu.cn; wangwei407@hrbeu.edu.cn).
			
	}}
	\markboth{Journal of \LaTeX\ Class Files, 2025}%
	{How to Use the IEEEtran \LaTeX \ Templates}
	\maketitle
	
	\begin{abstract}
		
		In this paper, we construct an air-sea collaborative system framework based on the Integrated Sensing and Communication (ISAC) techniques, where the Unmanned Aerial Vehicle (UAV) and Unmanned Surface Vehicle (USV) jointly inspect targets of interest while keeping communication with each other simultaneously. 
		First, we demonstrate the unique challenges encountered in this collaborative system, i.e., the coupling and heterogeneity of the UAV/USV's trajectories. Then, we formulate a total energy consumption minimization problem to jointly optimize the trajectories, flying and hovering times, target scheduling, and beamformers under the constraints of water currents, collision avoidance, and Sensing and Communication (S\&C) requirements. To address the strong coupling of the variables, we divide the original problem into two subproblems, namely, the hover point selection and the joint trajectory planning and beamforming design. 
		In the first subproblem, we propose a three-step hierarchical method including: (1) a virtual base station coverage (VBSC) and clustering algorithm to obtain the target scheduling and rough position of hover points; (2) a Bi-traveling salesman problem with neighborhood (Bi-TSPN)-based algorithm to determine the visiting order sequence of the hover points; (3) a hover point refinement and time allocation algorithm to further optimize the time allocation. 
		In the latter subproblem, we complete the remaining trajectory planning and beamforming design in each flying and hovering stage by developing a semi-definite relaxation (SDR) and successive convex approximation (SCA) method. Finally, we conduct a series of simulations to demonstrate the superiority of the proposed scheme over existing sequential access and leader-follower strategies.
		
	\end{abstract}
	
	
	\section{Introduction}

	\IEEEPARstart{W}ITH {the rapid development of the low-altitude economy, unmanned aerial vehicles (UAVs) have been widely used in maritime inspection due to their favorable attributes, including high mobility and flexibility, low cost, and high-resolution capabilities \cite{11072035,10879807}. Despite this, UAVs are constrained by limited energy and data storage capacity \cite{10815625}, severely restricting their use in relatively long-duration applications. To compensate for these shortcomings, an air-sea collaborative system is expected to be a promising solution in the maritime inspection field.}  
	
	{The collaborative operations of UAVs and unmanned surface vehicles (USVs) enable complementary advantages, significantly enhancing the effectiveness of tasks. Specifically, the UAVs with favorable visibility detect the targets, such as buoys, and capture the high-quality observations of interest. The USV can provide charging and edge computing support, enabling extended detection coverage and executing surface operations. Meanwhile, the UAVs communicate with the USVs to convey the data, including sensory results for data fusion, instructions for collaborative control, and data offloaded from targets or users \cite{10418158}. Therefore, the UAV-USV collaborative system is a typical use case that requires both sensing and communication (S\&C) functionalities.
		
		Recently, the emergence of integrated sensing and communication (ISAC) allows S\&C functions to share hardware and spectrum, which is expected to substantially improve the energy efficiency of systems \cite{10872967,10845869}. These advantages are ideally suited to miniaturized unmanned platforms, which motivates us to study an ISAC-empowered UAV-USV collaborative system, primarily focusing on trajectory and beamforming design. Compared to the existing works in the literature, two novel and interesting challenges have emerged in ISAC-empowered UAV-USV joint trajectory planning. 
		
		(1) \textit{\textbf{Coupling of trajectories}}: The trajectories of the UAV and USV are strongly coupled due to the S\&C task requirements mentioned above. For instance, UAV needs to fly close to the targets for sensing, while USV must maintain a certain distance from the UAV to ensure communication stability.
		
		(2) \textit{\textbf{Heterogeneity of trajectories}}: The UAV flies in free space at relatively high speed, while the motion of the slow-moving USV is greatly affected by water currents and obstacles. As a result, waiting time and speed control can significantly impact energy efficiency. }
	
	\subsection{ISAC-enabled UAV Systems}
	
	{
		The ISAC-based UAV systems have garnered extensive attention in recent years due to their high hardware integration and seamless S\&C capabilities \cite{10566041}. Existing studies primarily focus on UAVs as mobile S\&C base stations or relays in air-ground or air-sea environments, serving users such as vehicles and pedestrians \cite{10752639,10499863}. Furthermore, the purpose of system design for a single UAV base station primarily falls in the trajectory design and beamforming.
		
		In trajectory design, sensing performance (e.g., sensing signal-to-noise ratio (SNR) and mutual information), communication performance (e.g., transmission rate and bit error rate), and the energy consumption of the UAV are three key metrics of concern. Typically, one of them is designated as the optimization objective, while the others are imposed as constraints \cite{10529184,10295964,10100680}. Furthermore, the signal model and trajectory planning are also dependent on the various work modes. For instance, in many works, the UAV can perform S\&C tasks during flight \cite{9847217}. Although this work mode may save the entire time consumption, the Doppler shift induced by motion in flying reduces sensing effectiveness, requiring compensation algorithms and increasing operational complexity \cite{10713326}. Some other works adopt the flying and hovering mode, where the sensing task is only conducted when the UAV hovers over the targets \cite{10680299}. In the entire process, UAVs typically maintain a certain communication rate with users or base stations to guarantee a stable link. As for the beamforming design, the UAV typically performs sensing beamforming for multiple targets and communication beamforming for multiple users simultaneously \cite{10100680}. The purpose of beamforming is to form a directional beam and eliminate the interference between sensing targets, communication users, and S\&C services \cite{10769423}.
		
		However, the trajectory-planning problem with the consideration of cooperative slow-speed USV will be much more complicated due to the aforementioned challenges of \textit{coupling } and \textit{heterogeneity of trajectories}. Accordingly, beamforming and resource allocation require adaptive adjustment in response to the relative motion of the UAV and USV. This ISAC-enabled UAV-USV cooperative trajectory planning and beamforming design remains largely unexplored. 
		
	}

	\begin{figure}[t]
		\centering
		\includegraphics[width=8.5cm]{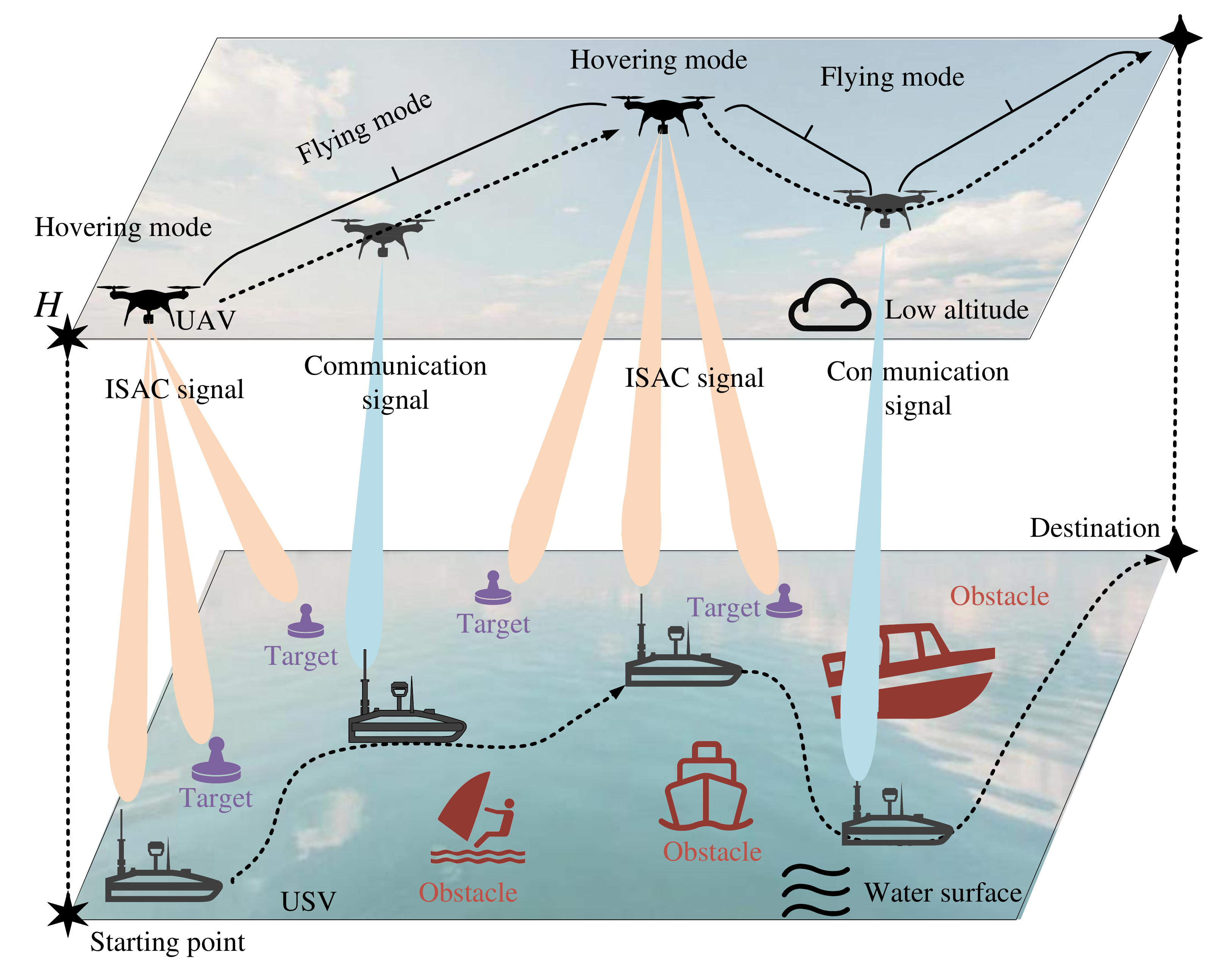}
		\caption{Maritime inspection system.}
		\label{fig0}
	\end{figure}
	
	\subsection{Maritime UAV-USV Cooperative Systems}

	Maritime cooperative unmanned systems are generally classified into two types: homogeneous and heterogeneous. In homogeneous maritime USV systems, research has primarily focused on achieving state synchronization and swarm coordination of multiple USVs through path planning and strategy design, while ensuring safe navigation under complex sea conditions \cite{10418158, 10787434}.
	Typical collaboration strategies include formation, leader–follower, and distributed schemes, mainly focusing on the priority of each individual in path planning \cite{9944188}. In contrast, current heterogeneous air–sea systems (UAV–USV teams) emphasize motion-control policies for cooperative search and pursuit, largely driven by the speed and endurance disparities between platforms \cite{10643681}. A prevalent approach is leader–follower, where the UAV first plans a path to rapidly detect targets and relay information, followed by a USV tracking this path \cite{10530448}. However, this framework ignores the coupling of UAV-USV trajectory caused by real-time S\&C requirements, and hence it is no longer capable in the ISAC-empowered air-sea cooperative system. In addition, jointly optimizing beamforming and trajectories promises gains in tracking accuracy, latency, and energy efficiency \cite{9085942,10423261}.
	Therefore, compared with the existing works, the ISAC-enabled UAV-USV cooperative framework still faces two major challenges: (1) breaking through the traditional leader-follower paradigm to explore more efficient joint trajectory design methods, and (2) seeking a balance between S\&C within the cooperative framework while approaching performance limits.
	
	\subsection{Contributions}
	In this paper, we propose an ISAC-empowered UAV-USV collaborative framework in maritime inspection, where the total energy consumption is minimized while meeting the cumulative sensing SNR and communication rate requirements. Inspired by \cite{10680299}, we adopt a multi-stage S\&C process including \emph{hovering mode} and \emph{flying mode}.
	
	As shown in Fig.~\ref{fig0}, during each flying and hovering stage, the UAV first flies to the next hover point while maintaining communication with the USV, and then performs S\&C tasks once it arrives.
	Thus, the complicated trajectory and beamforming design problem can be transformed into the following two sub-problems.
	
	(1) \textbf{\textit{Hover point selection}}:\textit{ Given the random distribution of sensing targets, how can we determine the number, the positions of the hover points, and the UAV hover time at each point to minimize the total energy consumption?}
	
	(2) \textbf{\textit{Joint trajectory planning and beamforming}}: \textit{Given the hover points, how can we jointly design UAV–USV trajectories and beamforming by considering water currents and obstacles to satisfy S\&C requirements? }
	
	In summary, our main contributions are as follows:
	
	\begin{itemize}[]
		\item First, we establish a multiple flying and hovering stages UAV-USV collaborative framework, where the original energy minimization problem is transformed into the hover points selection problem from start to end-point, and the joint trajectory planning and beamforming problem at each stage. Thus, the complicated joint trajectory planning with strongly coupled variables such as UAV and USV positions, hovering duration, visiting order, and resource allocation becomes tractable.    	
		\item 
		Second, to address the hover point selection problem, we propose a three-step hierarchical method including: (1) a virtual base station coverage (VBSC) and clustering algorithm to obtain the target scheduling and rough position of hover points; (2) a Bi-traveling salesman problem with neighborhood (Bi-TSPN)-based algorithm to determine the visiting order sequence of the hover points; (3) a hover point refinement and time allocation algorithm to further optimize the time allocation. 
		
		\item
		Third, we establish an energy minimization model for both flying and hovering modes under complex environments with obstacles and water currents. The model jointly optimizes the trajectories of the UAV and USV, as well as the UAV’s S\&C beams. 
To solve this problem, we propose an alternating optimization algorithm that iteratively optimizes both the trajectories and the beams.
		\item 
		
		Finally, the experimental results demonstrate that:
		(1) The proposed method significantly reduces energy consumption compared to sequential access and leader–follower strategies.
		(2) With the same number of targets, a more dispersed target distribution leads to a higher rate of energy consumption growth.
		(3) A higher SNR notably increases energy consumption, highlighting the trade-off between performance and efficiency.

	\end{itemize}
	
	\textbf{\textit{Notations:}} The uppercase bold letter $\mathbf{A}$, the lowercase bold letter $\mathbf{a}$, the normal letter $a$, and the fraktur letter $\mathcal{A}$ denote a matrix, a vector, a scalar, and a set, respectively. $\| \cdot \|$ denote the Euclidean norm. $\mathbf{A} \succeq$ 0 means that $\mathbf{A}$ is positive semi-definite. $\mathrm{rank}(\mathbf{A})$ and $\mathrm{tr}(\mathbf{A})$ denote the rank and trace of matrix $\mathbf{A}$, respectively. $\mathbb{C}$ denotes the complex space. $\mathbf{a}^H$ denotes the Hermitian (conjugate transpose) of the vector $\mathbf{a}$.
	$\mathbf{I}_M\in \mathbb{R}^{M \times M}$ is an unit matrix.
	$\mathbb{E}
	(\cdot)$ for the stochastic expectation.
	
	\section{System Model}
	\subsection{Signal Frame Structure}
	\label{Sec:2}
	\label{Sec:3}

	In this paper, we consider a quadrotor UAV equipped with $M$ antennas to sense $K_\text{tar}$ randomly distributed targets, followed by a slow-speed USV that supports data offload, edge computing, etc., as shown in Fig. \ref{fig0}. In addition, $K_\text{obs}$ obstacles are distributed on the water surface. The flying and hovering mode in \cite{10680299} is adopted, where the entire stage, from start to endpoint, is divided into several hover points. Specifically, the UAV must maintain a stable communication link with the USV throughout the entire stage and implement the sensing tasks only in hovering mode \footnote{ 
		Sensing in hovering mode can avoid the impact of Doppler effects caused by motion maneuvers\cite{10680299}.
	}. 
	After inspecting all $K_\text{tar}$ targets, the UAV and USV reach the destination at the same time. In this process, the trajectories of UAV and USV and the S\&C performance are strongly coupled, which motivates us to establish an efficient UAV-USV collaborative framework.

	\begin{figure}[!t]
		\centering
		\includegraphics[width=9cm]{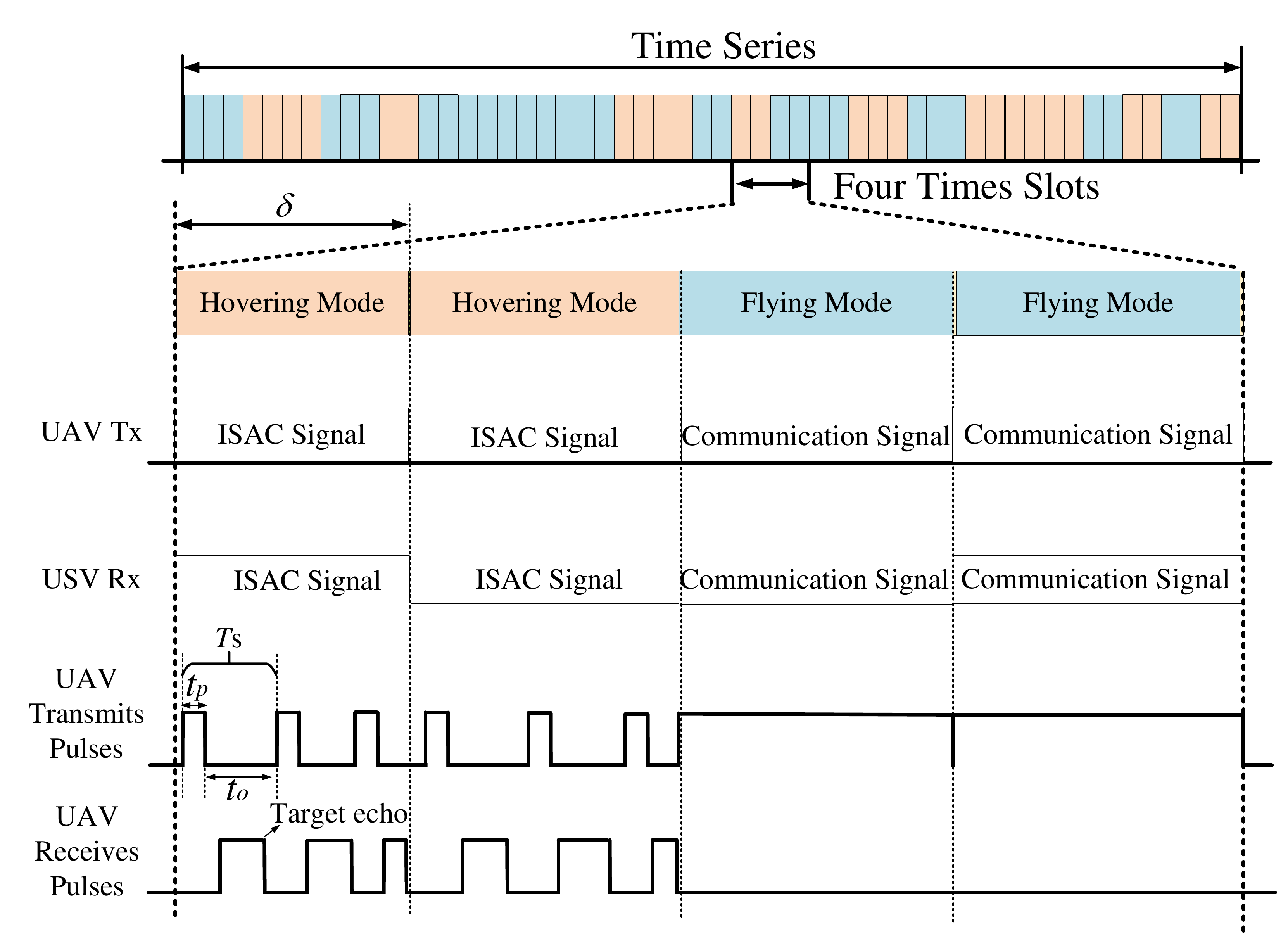}
		\caption{ISAC time slot structure.}
		\label{nfig2}
	\end{figure}
	
	The detailed frame structure is illustrated in Fig. \ref{nfig2}. 
	The total runtime $T$ is divided into $\mathcal{N}$ equal time slots with each of duration $\delta$.
	We denote the index set $\mathcal{N} = \{1,2,\dots,N\}$, the flying mode index set $\mathcal{F}$, and the hovering mode index set $\mathcal{H}$ that satisfies $\mathcal{F} \cap \mathcal{H} = \varnothing, \mathcal{F} \cup \mathcal{H} = \mathcal{N}$. For the sensing task, each time slot is further divided by $N_s$ scanning rounds with each duration of $T_s=\frac{\delta}{N_s}=t_p+t_o$. Specifically, in each round, the UAV first transmits a scanning pulse of duration $t_p$, then immediately switches to a listening mode to receive the echo corresponding to that pulse, with a listening time of $t_o$.
	
	We denote $\mathbf{t}_k = (t_{x_k}, t_{y_k}, 0)^T $ and $ \tilde{\mathbf{t}}_{k'} = ( \tilde{t}_{x_{k'}}, \tilde{t}_{y_{k'}}, 0)^T $ as the positions of the $k$-th target and $k'$-th obstacle, respectively. $\mathbf{q}[n] = (q_x[n], q_y[n], H)^T$ and $\mathbf{b}[n] = (b_x[n], b_y[n], 0)^T$ are the trajectory points of the UAV and USV at $n$-th time slot \footnote{ In this paper, we focuse on 2D motion where the UAV flies at a fixed altitude $H$. Furthermore, each time slot is small enough for the UAV and USV positions to remain approximately constant. }. 
	Moreover, we define the sensing flag \( r_k[n] \in\{0,1\}\) to indicate whether the \( k \)-th target is sensed at time slot \( n \), i.e., $r_k[n]=1$ only if the target is sensed.

	\subsection{Communication and Sensing Channel Model}
	
	\textit{1) Communication Channel:}
	The air-sea channel model from UAV to USV can be expressed by \cite{9453748}
	\begin{equation}
		\begin{aligned}
			\mathbf{h}_{c}[n] &  =\frac{\rho_0\iota}{d^{2}_{\text{com}}[n] }\mathbf{a}(\mathbf{q}[n],\mathbf{b}[n]),
			\label{Ch}
		\end{aligned}
	\end{equation}
	where \( \frac{\rho_0}{d_{\text{com}}^2[n]} \) represents the large-scale fading coefficients, with \( \rho_0 \) being the channel gain. \( d_\text{com}[n] = \|\mathbf{q}[n] - \mathbf{b}[n]\| \) is the Euclidean distance between the UAV and the USV.
	\( \iota \) is the small-scale fading coefficients.
	The array response vector of uniform linear array (ULA) can be given by
	\begin{equation}
		\mathbf{a}( \mathbf{q}[n],\mathbf{b}[n])=[1,
		e^{j2\pi d\frac{\cos(\varphi[n])}{\lambda}},
		\ldots,e^{j2\pi(M-1)d\frac{\cos(\varphi[n])}{\lambda}}]^T,
	\end{equation}
	where $d$ and $\lambda$ are the antenna space and wave length, respectively, and $\varphi[n] =\arccos\frac{H}{ \|\mathbf{q}[n]-\mathbf{b}[n]\|    }$.

	\textit{2) Sensing Channel:}
	The round-trip sensing channel between the UAV and the $k$-th target can be expressed as \cite{liu2023fair}
	\begin{equation}
		\mathbf{H}_{k}[n]=\frac{\beta\epsilon_k[n]}{2d_{\text{sen},k}[n]} \mathbf{a}(\mathbf{q}[n],\mathbf{t}_k)\mathbf{a}^H(\mathbf{q}[n],\mathbf{t}_k),
	\end{equation}
	where $\beta$, $\eta$ represent the channel power gain and the mean radar cross section, respectively. $d_{\text{sen},k}[n]= \|\mathbf{q}[n]-\mathbf{t}_k\|$ is the distance from the UAV to the $k$-th target, and $\epsilon_k[n] =  \sqrt{\frac{\eta}{4\pi d^2_{\text{sen},k}[n]}}$ is the reflection coefficient of the target $k$.

	\subsection{Signal Model}
	
	\textit{1) Flying Mode:}
	In this mode, the UAV flies from one hover point to the next, while communicating with the USV by transmitting the communication-only signals. The signal model can be written by
	\begin{equation}
		\begin{aligned}
			\mathbf{x}_f[n,n_s]=&\sqrt{\frac{t_p}{\delta}}\mathbf{w}_f[n]{c}_f[n,n_s],\: n \in \mathcal{F}, n_s = 1,...,N_s,\\
		\end{aligned}
	\end{equation}
	where 
	$\mathbf{w}_f[n] \in\mathbb{C}^{M\times1}$  is the beamforming
	vector. 
	$c_f[n,n_s]$ denotes the downlink communication signal transmitted from the UAV to the USV at time slot $n$ during the $n_s$-th round.
	Thus, the received signal at the USV can be given by
	\begin{equation}\begin{aligned}
			{y}_\text{usv}[n,n_s] & ={\mathbf{h}^H_{c}[n] \mathbf{x}_f[n,n_s] }+{{z}_f[n,n_s]},
	\end{aligned}\end{equation}
	where ${z}_f[n,n_s]\in\mathcal{CN}(0,\sigma_c^2)$ represents additive White Gaussian noise. 
	Here, $\mathbf{h}_{c}[n]$ and $\mathbf{w}_f[n]$ are independent of $n_s$ as the channel remains constant within a time slot. The sum SNR at time slot $n$ can be computed as
	\begin{equation}
		\begin{aligned}
			\gamma_f[n]  = &\sum_{n_s =1}^{N_s}\gamma_f[n,n_s] = N_s {\frac{t_p}{\delta}}\frac{
				\|\mathbf{h}^H_{c}[n] \mathbf{w}_f[n]\|^2  }{ \sigma_c^2 }.
			\label{ccc} 
		\end{aligned}
	\end{equation}
	Therefore, the data transmission rate can be given by 
	\begin{equation}
		{R}_f[n]=\log_2\left(1+\gamma_f[n]\right).
		\label{com}
	\end{equation}
		
		\textit{2) Hovering Mode:}
		In this mode, the UAV employs an ISAC signal to maintain communication with the USV while sensing the targets simultaneously \cite{9124713}. The ISAC signal transmission model is written as
		\begin{equation}
			\begin{aligned}
				\mathbf{x}_h[n,n_s]=&\sqrt{\frac{t_p}{\delta}}(
				\sum_{k=1}^{K}{r}_{k}[n]\mathbf{v}_{{k}}[n] {{s}_k}[n,n_s]
				+\mathbf{w}_h[n]\\
				&\times{c}_h[n,n_s]),\:\: n \in \mathcal{H}, n_s = 1,2,...,N_s,
			\end{aligned}
		\end{equation}
		where
		$\mathbf{w}_h[n] \in \mathbb{C}^{M \times 1}$ and $\mathbf{v}_k[n] \in \mathbb{C}^{M \times 1}$ denote
		communication beamforming vector for the USV and the sensing beamforming vector for the $k$-th target, respectively. $c_h[n,n_s]$ and $s_k[n,n_s]$ are the corresponding communication and sensing signals.
		We assume that the S\&C signals are zero-mean, temporally white, wide-sense stationary, satisfying 
		\begin{equation}
			\mathbb{E}
			({s}_k[n,n_s]{c}^H_h[n,n_s])={0}.
		\end{equation}

		\textit{Communication Received Signal Model:} The signal received by the USV can be expressed by
		\begin{equation}
			\begin{aligned}
				& \tilde{y}_{\text{usv}}[n,n_s] = \sqrt{\frac{t_p}{\delta}} \underbrace{\mathbf{h}^H_{c}[n] \mathbf{w}_h[n]{c}_h[n,n_s]}_{\text{Intended signal}}\\ &+\sqrt{\frac{t_p}{\delta}}\underbrace{\sum_{k=1}^{K}{r}_{k}[n]\mathbf{h}^H_{c}[n]\mathbf{v}_{{k}}[n]{{s}_k[n,n_s]}}_{\text{Sensing interference}}+{{z}_h[n,n_s]},
			\end{aligned}
			\label{new111}
		\end{equation}
		where ${{z}_h[n,n_s]} \in\mathcal{CN}(0,\sigma_h^2 ) $ represents additive Gaussian noise.
		The sum SNR of downlink communication for USV can be computed by
		\begin{equation}
			\begin{aligned}
				\gamma_h[n] & ={
					\frac{N_s {\frac{t_p}{\delta}}\|\mathbf{h}_{c}^H[n] \mathbf{w}_h[n]\|^2}{N_s \frac{t_p}{\delta}  \sum_{k=1}^{K}  {r}_{k}[n]\|\mathbf{h}_{c}^H[n]\mathbf{v}_{{k}}[n]\|^{2}+\sigma_h^2  }	  }.
			\end{aligned}
		\end{equation}
		The communication rate is given by 
		\begin{equation}
			{R}_h[n]=\log_2\left(1+\gamma_h[n]\right).
			\label{zhuanhuanrc}
		\end{equation}
		
		
		\textit{Sensing Received Signal Model:} The echo signal of the $k$-th target collected by the UAV can be expressed as 
		\footnote{  Communication interference is ignored since the UAV can extract the component of the sensing signal by subtracting the known downlink communication and decoded USV signals \cite{10100680}.} 
		\begin{equation}
			\begin{aligned}
				&{y}_k[n,n_s] = \sqrt{\frac{t_p}{\delta}}( \underbrace{{{\mathbf{u}}}^H_k[n]{ \mathbf{H}_{k}[n]{{\mathbf{v}}}_{{k}}[n]{s}_k[n,n_s] }}_{\text{Intended signal}}\\ +&\underbrace{\sum_{j=1,j\neq k}^{K}{r}_{j}[n]{{\mathbf{u}}}^H_k[n]{ \mathbf{H}_{k}[n]{{\mathbf{v}}}_{{j}}[n]{s}_j[n,n_s]) }}_{\text{Sensing interference}}+{{\mathbf{u}}}^H_k[n]{\mathbf{z}_k[n,n_s]},
			\end{aligned}
		\end{equation}
		where $\mathbf{z}_{k}[n,n_s] \in\mathcal{CN}(0,\sigma_s^2 \mathbf{I}_M ) $ and ${\mathbf{u}}_k[n]$ represent the noise and the receive combining vector for $k$-th target, respectively.
		Thus, the sensing SNR at time slot $n$ for the $k$-th target can be expressed by
		\begin{equation}
			\gamma_k[n]=\frac{{\frac{N_st_p}{\delta}}\| {{{\mathbf{u}}}}_k^H[n]{ \mathbf{H}_{k}[n]\mathbf{v}_{{k}}[n] } \|^2 }{ {\sum_{j\neq k}^{}{r}_{j}[n]\frac{N_st_p}{\delta}}\|{{{\mathbf{u}}}}_k^H[n]
				{ \mathbf{H}_{k}[n]\mathbf{v}_{{j}}[n] }\|^2+\sigma_s^2\|{{\mathbf{u}}}_k[n]\|^2}.
			\label{sen121}
		\end{equation}
		
		
		\subsection{{Energy Consumption Model}}
		
		The UAV's power consumption is highly dependent on its velocity, defined by $ \mathbf{v}_{\text{uav}}[n] = ( \mathbf{q}[n] - \mathbf{q}[n-1]) / \delta $. In hovering mode, that is, the speed of $ v_\text{uav}[n] = \| \mathbf{v}_{\text{uav}}[n] \|=0$, power consumption remains constant as $p_\text{uav}^h[n] = U_\text{uav}^0 + U_\text{uav}^1,\: n \in \mathcal{H}$, where $U_\text{uav}^0$ and $U_\text{uav}^1$ are the profile power and induced power, respectively. In flying mode, the power consumption is \cite{10100680} 
		\begin{equation}
			\begin{aligned}
				p_\text{uav}^f[n]  &=U_\text{uav}^0\left(1+\frac{3{v}^2_\text{uav}[n]}{{U_{\text{tip}}}^2}\right) +\frac{1}{2}d_0\rho \varphi A{v}^3_\text{uav}[n]\\
				& +U_\text{uav}^1\left(\sqrt{\left(1+\frac{{v}^4_\text{uav}[n]}{4{v_0}^4}\right)}-\frac{{v}^2_\text{uav}[n]}{2{v_0}^2}\right)^{1/2},\: n \in \mathcal{F}.\label{fly11}
			\end{aligned}
		\end{equation}
		Here, $U_{\text{tip}}$ is the tip speed of the rotor blade, $v_0$ is the mean induced speed of the rotor during forward flight, $d_0$ is the fuselage drag coefficient, $\rho$ is the air density, $\varphi$ is the rotor solidity, and $A$ is the rotor disc area. Therefore, the total energy consumption of the UAV can be given by 
		\begin{equation}
			\begin{aligned}
				&E^{\text{total}}_{\text{uav}} = \underbrace{\sum_{n=1}^{N} 
					\mathbb{1}( n \in \mathcal{F} )( p_\text{uav}^f[n]+\|\mathbf{w}_{f}[n]\|^2)}_{\text{ Energy consumption in flying mode
				}}\\
				&
				+
				\underbrace{\sum_{n=1}^{N} \mathbb{1}( n \in \mathcal{H} ) (p_\text{uav}^h[n]
					+\|\mathbf{w}_{{h}}[n]\|^2
					+\sum_{k=1}^{K}{r}_k[n]\|\mathbf{v}_{{k}}[n]\|^2)}_{\text{ Energy consumption in hovering mode
				}},
			\end{aligned}
		\end{equation}
		where $\mathbb{1}(\cdot)$ represents indicator function.
		
		Similarly, let $\mathbf{v}_\text{usv}[n] = ({ \mathbf{b}[n]} -{ \mathbf{b}[n-1]}) /  \delta$ denote the velocity of USV, then the total energy consumption of USV can be expressed as \cite{1315730}
		\begin{equation}
			\begin{aligned}
				&E^{\text{total}}_{\text{usv}}=\sum_{n=1}^{N}\alpha \|\mathbf{v}_\text{usv}[n]- \mathbf{v}_{w}[n]   \|^2,\label{h111}
			\end{aligned}
		\end{equation}
		where $
		\mathbf{v}_{w}[n] = \left( v_{x,\mathbf{b}[n]}, v_{y,\mathbf{b}[n]}, 0 \right)^{T}$ is the water current velocity at the point $\mathbf{b}[n]$, and constant $\alpha$ depends on the size of the USV and the properties of water \footnote{ 
			The flow data can be obtained from historical hydrological records or measured in real time by the flow sensors onboard the USV.
		}.
		
		\subsection{Problem Formulation}
		
		Our aim is to minimize the total energy consumption of UAV and USV while meeting the S\&C requirements. However, unlike existing works on ISAC-based UAV trajectory planning, the collaborative system encounters a novel challenge due to the \textit{{heterogeneity of UAV and USV}}. Specifically, in the literature \cite{10752639,10499863,10680299,9847217,10295964}, the total run-time or hover-time (or both) is always assumed to be fixed, which is unreasonable in our scheme. The difference in the velocities of the UAV and USV restricts them from waiting for each other at some point to ensure the stability of the communication link \footnote{Communication quality depends on the distance between UAV and USV.}. Therefore, the total time interval $N$ and the hover time slots set $\mathcal{H}$ must be treated as variables to be determined. This significantly complicates the trajectory planning problem.     
		
		Furthermore, the trajectories ($\mathbf{q}[n],\mathbf{b}[n]$), the beamformers for S\&C ($\mathbf{w}_{f}[n],\mathbf{w}_{h}[n],\mathbf{v}_{k}[n]$), the target indices to be sensed at each hover point ($r_k[n]$) also dramatically influence the collaborative performance. Let us denote the variables as $\Theta = \{
		N,\mathcal{H}, \mathbf{q}[n],\mathbf{b}[n],\mathbf{w}_{f}[n],\mathbf{w}_{h}[n],\mathbf{v}_{k}[n],r_k[n]
		\}$, then the optimization problem can be formulated by
		\begin{subequations}
			\begin{align}
				\mathrm{(P1):}&\min_{
					\Theta
				} E^{\text{total}}_{\text{uav}}+E^{\text{total}}_{\text{usv}}\notag \\
				\mathrm{s.t.}\quad
				&	{R}_f[n]\geq\Gamma_f,   n \in \mathcal{F}; \:\:	{R}_h[n]\geq\Gamma_h,  \: n \in \mathcal{H},
				\label{p0-1}\\
				&			\sum_{n=1}^{N}\gamma_k[n]\geq\Gamma_s^{\text{total}},  \: k = {1,...,K_\text{tar}}, \: n \in \mathcal{H},
				\label{p0-9}\\
				&	0\leq \|   \mathbf{w}_{{f}}[n]\|^2\leq p_{\text{max}}, \: n \in \mathcal{F},
				\label{p0-8}\\
				&\sum_{k=1}^{K}r_{k}[n]\|\mathbf{v}_{{k}}[n]\|^2+\|\mathbf{w}_h[n]\|^2\leq p_{\text{max}}, \:n \in \mathcal{H},	\label{p0-10}\\
				&r_k[n]=\{0,1\}, \:n \in \mathcal{H},	\label{p0-11}\\
				&	\| \mathbf{b}[n]-\tilde{\mathbf{t}}_{k'}\|\geq r_1,
				\:\: n \in \mathcal{N}, \: k' = {1,...,K_\text{obs}}, 
				\label{p0-3} \\ 
				& \|\mathbf{v}_\textbf{uav} [n]\| \leq v_{\text{uav}}^{\text{max}},  \: \:  \|\mathbf{v}_\textbf{usv} [n]\| \leq v_{\text{usv}}^{\text{max}}, \: n \in \mathcal{N}.   \label{p18h}       
			\end{align}
		\end{subequations}
		
		Constraint (\ref{p0-1}) and (\ref{p0-9}) are the communication and sensing requirements, respectively. Particularly, we restrict a minimum rate ($\Gamma_f$ for flying and $\Gamma_h$ for hovering) in each time slot for communication and an accumulated SNR for each target ($\Gamma_s^{\text{total}}$). 
		Constraints (\ref{p0-8}-\ref{p0-10}) are the maximum power budget at the UAV. The constraint (\ref{p0-3}) ensures that the USV trajectory avoids collisions with any obstacles, and \eqref{p18h} represents the velocity limitations. In addition, we imply that the start and end points of $\mathbf{q}[n]$ and $\mathbf{b}[n]$ are given initially according to the specific inspection task. 
		
		It should be highlighted that solving the original problem $\mathrm{P1}$ directly is quite challenging, as all variables are strongly coupled by introducing the time variable $N$ and the hover time set $\mathcal{H}$. This motivates us to develop an efficient approximation method to obtain a suboptimal solution with tolerable performance loss. In this hovering and flying mode, the hover points and dwell times in each hover point play a critical role in the entire trajectory planning. Once hover points and dwell times are determined, the trajectory and beamformer can be designed individually in each stage between two adjacent hover points, which significantly simplifies the original problem. Consequently, we transform problem ($\mathrm{P1}$) into the following two subproblems:
		
		(1) \textbf{\textit{Hover point selection}}: 
		With the goal of minimizing energy consumption and considering S\&C performance, we determine the number and locations of hover points, the duration of each flying and hovering mode.
		
		(2) \textbf{\textit{Joint trajectory planning and beamforming}}: 
		With the goal of minimizing energy consumption and considering water currents and obstacles in each stage, we jointly optimize the trajectory and beamforming to meet the S\&C requirements. 
		
\begin{figure}[!t]
			\raggedright
			\includegraphics[width=8.7cm]{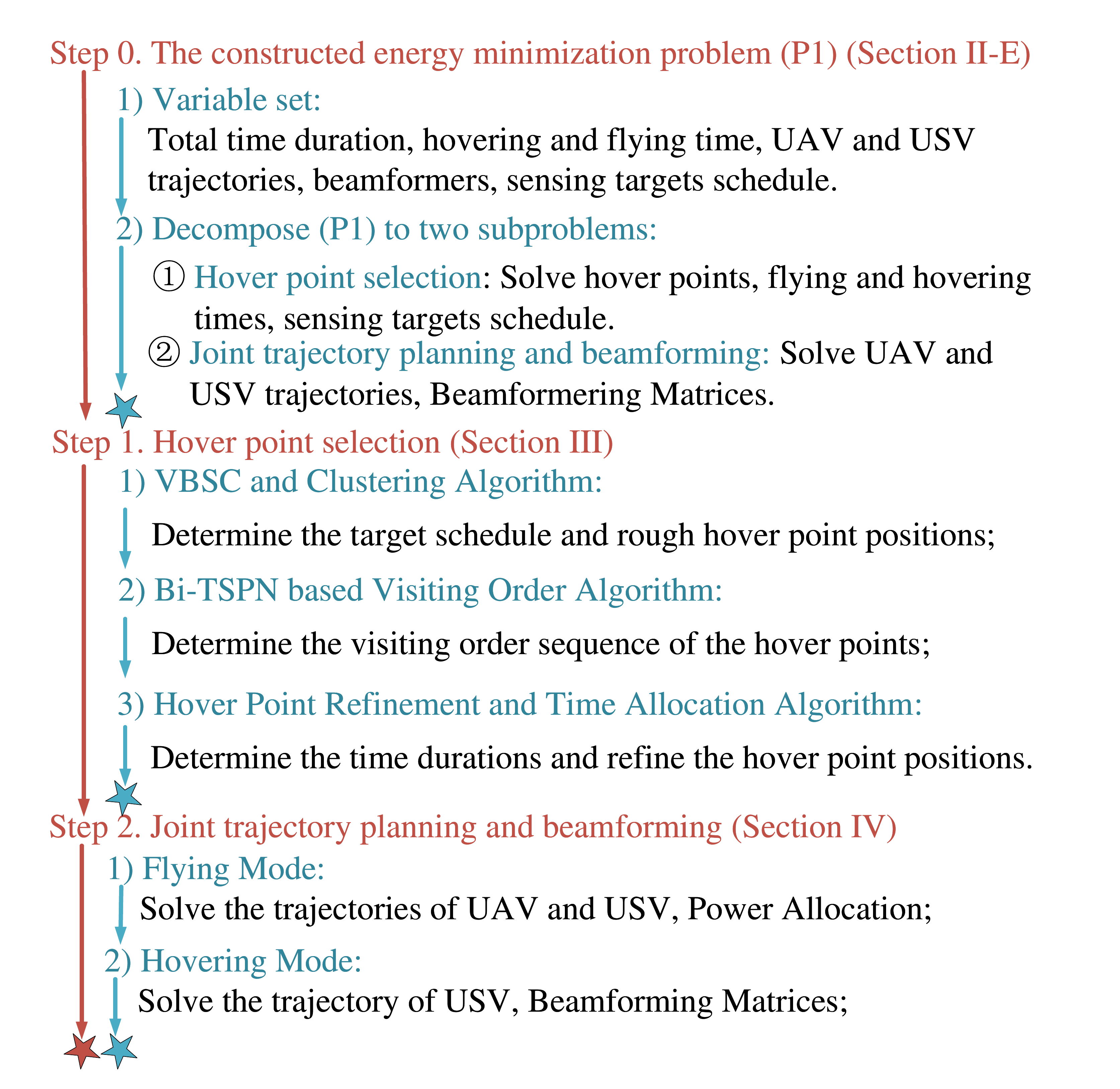}
			\caption{Flow of the problem solution.}
			\label{liucheng}
		\end{figure}

		To clearly illustrate the analytical process and research methods adopted in this study, an overall flowchart is provided to show the methods and optimization variables solved in each section, as shown in~Fig. \ref{liucheng}.

		\newtheorem{remark}{Remark}

		\label{3}

		\section{ {Hover Points Selection} }

		\subsection{Problem Analysis}

		\begin{figure}[!t]
			\centering
			\includegraphics[width=7.5cm]{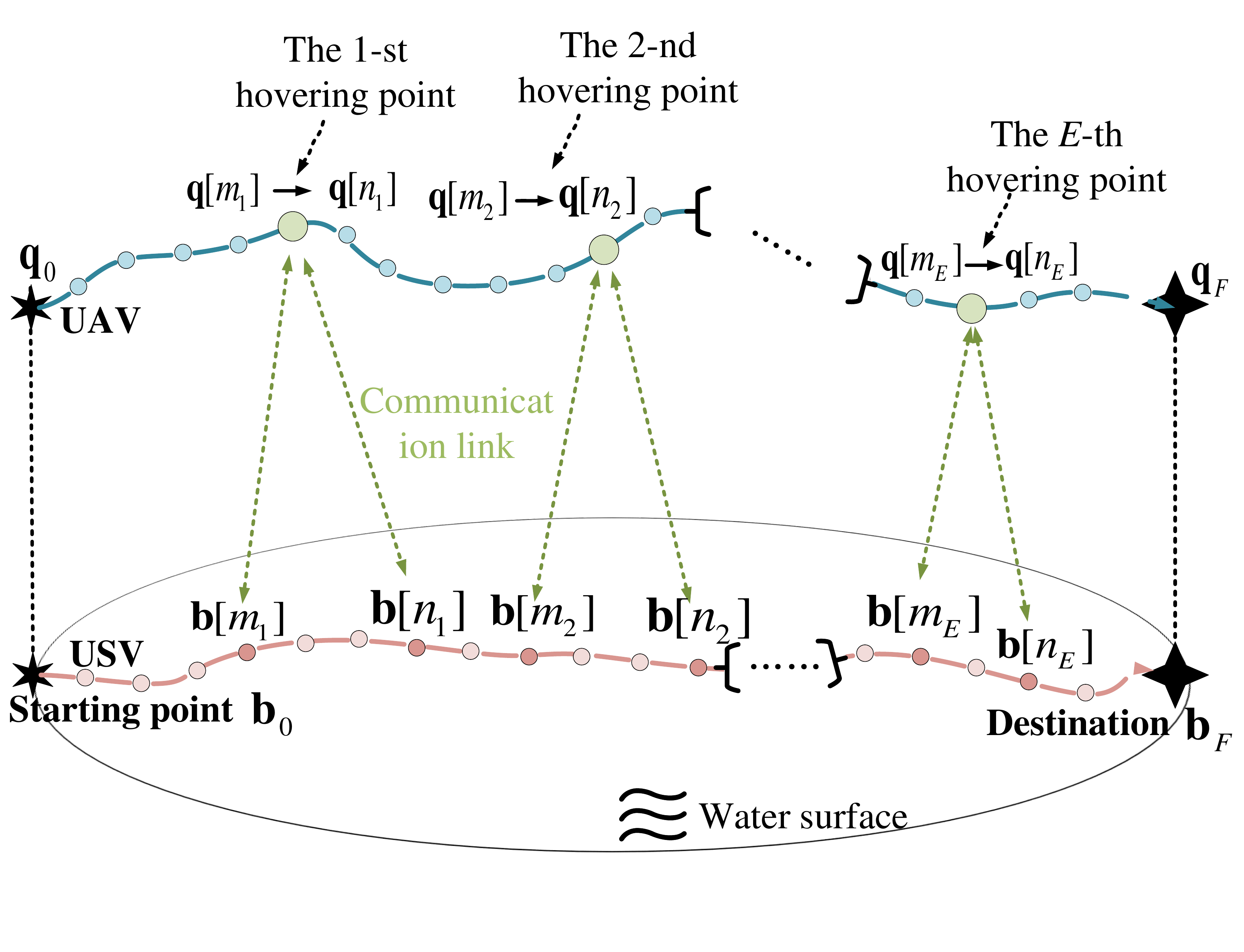}
			\caption{Diagram of UAV and USV trajectories in our flying and hovering scheme.}
			\label{time1}
		\end{figure}
		
		{ As illustrated in Fig. \ref{time1}, we assume that the entire time horizon is divided into $E+1$ stages by $E$ hover points. For the convenience of discussion, we define \(m_e, n_e \in \mathcal{H}\) as the start and end moment of the hovering mode in \( e \)-th stage. Thus, the duration of the $e$-th stage is $n_e-n_{e-1}+1$, with the hover duration being $n_e-m_e+1$. The purpose of this section is to determine $E$, $m_e$, $n_e$, and the position of the UAV hover point $\mathbf{q}[m_e]$, with $e = 1, \dots, E$.   
			
			Apparently, the selection of the hover point also depends on the other variables in $\Theta$. To proceed, we temporarily decouple these variables by making the following assumptions to obtain a coarse solution, which will be refined in the next section.       
			
			(A.I) The beamforming design is temporarily ignored, and we choose the Maximum Ratio Transmission (MRT) scheme for the S\&C beamformer as
			\begin{equation} \notag
				\mathbf{w}_f[n] = \mathbf{w}_h[n] = \frac{\sqrt{p_c} \mathbf{a}(\mathbf{q}[n], \mathbf{b}[n])}{\|\mathbf{a}(\mathbf{q}[n], \mathbf{b}[n])\|}, 
			\end{equation}
			\begin{equation} \notag
				\mathbf{v}_{k}[m_e] =  \frac{ \sqrt{p_s} \mathbf{a}(\mathbf{q}[m_e], \mathbf{t}_k)} {\|\mathbf{a} (\mathbf{q}[m_e], \mathbf{t}_k)\|}, \: \mathbf{u}_k[m_e] = \frac {\mathbf{a}(\mathbf{q}[m_e], \mathbf{t}_k)}{\|\mathbf{a}(\mathbf{q}[m_e], \mathbf{t}_k)\|}.
			\end{equation}
			
			\begin{figure*}[!t]
				\centering
				\includegraphics[width=18cm]{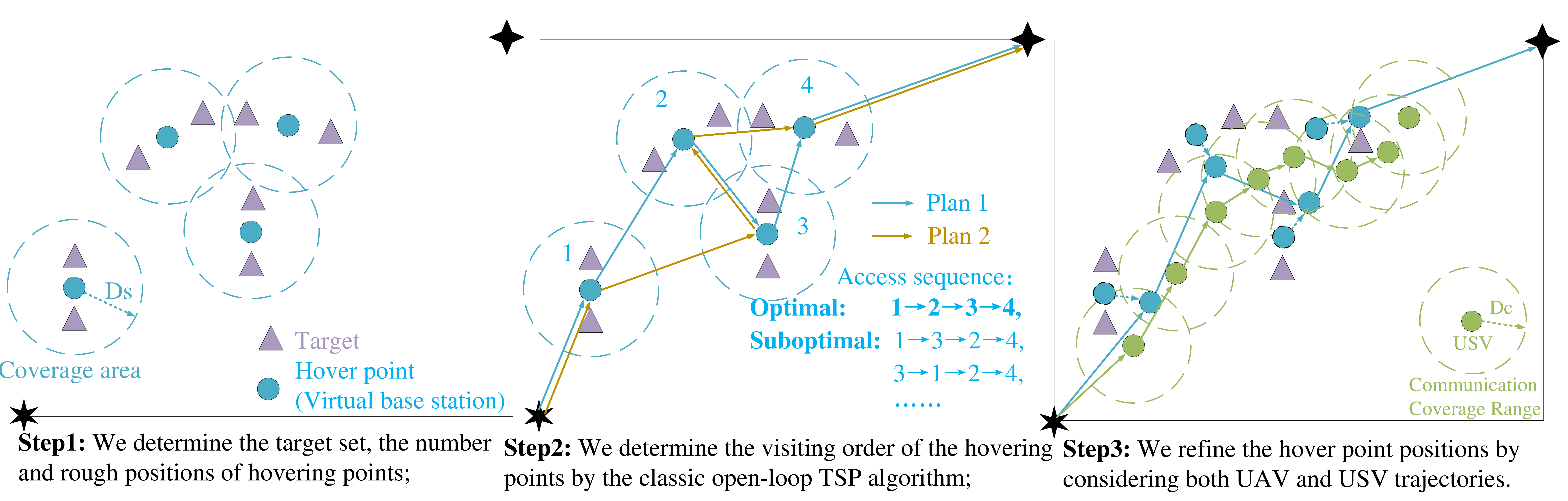}
				\caption{The diagram for the three-step hierarchical hover point selection method.}
				\label{time2}
			\end{figure*}
			
			(A.II) The power allocation between S\&C functions is temporarily ignored, and we adopt the constant power $p_s$ and $p_c$ for S\&C while evenly allocating the sensing power for each target to be sensed. 
			
			(A.III) The sensing interference in \eqref{new111} is temporarily neglected to eliminate the coupling between the communication rate and the selected targets, leaving interference cancellation by beamformer design in the next section.
			
			As demonstrated in \cite{8255824}, the hover point selection problem involves not only determining the hover point positions but also the visiting order of the point sequence, which resembles the traditional Traveling Salesman Problem (TSP) problem. For illustration convenience, we first transform the S\&C constraints into distance-related forms and seek the solution using a TSP-like approach.   
			
			Let us denote $\Gamma_f = \Gamma_h = \Gamma_c$, the communication constraint (\ref{p0-1}) can be recast by applying Assumption (A.I) as
			\begin{align}\label{comconstraint}
				\|\mathbf{q}[n]-{\mathbf{b}}[n]\|  \leq  \left( \frac{M p_c \rho_0^2 \iota^4}{\sigma_c^2 \left(2^{\Gamma_c} - 1\right)} \right)^{\frac{1}{4}} \triangleq    D_c(\Gamma_c,p_c).
			\end{align}
			It can be observed that the communication distance threshold $D_c(\Gamma_c,p_c)$ is proportional to the transmit power $p_c$ and inversely proportional to the rate requirement $\Gamma_c$. On the other hand, for sensing constraint \eqref{p0-9}, the cumulative SNR threshold $\Gamma_s^\text{total}$ depends on the dwell time (or hover time) and the instantaneous SNR at a single time slot defined by $\Gamma_s$. Similarly, by applying Assumption (A.I) to (\ref{sen121}) and introducing a new instantaneous SNR constraint at $e$-th hover point by 
			\begin{align}
				\gamma_k[m_e] =  N_s {\frac{t_p}{\delta}}{\frac{\eta\beta^2 p_{s,k}  M }
					{16\pi \sigma_s^2  {\|  \mathbf{q}[m_e]-\mathbf{t}_k \|}^4}}  \geq \Gamma_s,
			\end{align}
			where $p_{s,k} = {p_s}/{ \sum_{k=1}^{K}r_{k}[m_e]}$ according to the Assumption (A.II). After a simple mathematical transformation, we can determine the sensing distance for the $k$-th target, i.e.,
			\begin{align} \label{sensingconstraint}
				\|{\mathbf{q}}[m_e]-{\mathbf{t}}_k\|   \leq  \left(N_s \frac{t_p}{\delta} \frac{\eta \beta^2 p_{s,k} M}{16\pi \Gamma_s \sigma_s^2} \right)^{\frac{1}{4}} \triangleq    D_{s,k}(\Gamma_s,p_{s,k}).
			\end{align}
			The sensing distance threshold $D_{s,k}(\Gamma_s,p_{s,k})$ is proportional to the transmit power allocated for the $k$-th target $p_{s,k}$ and inversely proportional to the instantaneous SNR requirement $\Gamma_s$.  Furthermore, the hover time of the $e$-th point satisfies 
			\begin{align} \label{timeconstraint}
				n_e - m_e + 1 \ge \max_{k\in\mathcal{K}}\; r_{k}[m_e]\frac{\Gamma_s^{\text{total}}}{\gamma_k[m_e]}.  
			\end{align}
			Note that the right-hand side of the inequality includes the selection of the targets to be sensed at the $e$-th point.  
			
			For the objective in ($\mathrm{P1}$), we use the average speed in the $e$-th stage instead of the instantaneous speed in single time slot $n$. Concretely, the average speeds of the UAV and the USV in flying and hovering modes are given, respectively, by
			\begin{align} \notag
				\bar{v}_{\text{uav},e} = \| \mathbf{q}[m_e]- \mathbf{q}[n_{e-1}] \| /(m_e-n_{e-1}), \\ 
				\bar{v}^f_{\text{usv},e} = \| \mathbf{b}[m_e]- \mathbf{b}[n_{e-1}]\| / (m_e-n_{e-1}), \notag \\ 
				\bar{v}^h_{\text{usv},e} = \| \mathbf{b}[n_e]- \mathbf{b}[m_e]\| / (n_e-m_{e}). \notag
			\end{align}
			
			By substituting the average speed into (\ref{fly11}) and (\ref{h111}), and denoting the average power of the UAV and the USV in flying and hovering mode by $\bar{p}_{\text{uav},e}$, $\bar{p}^f_{\text{usv},e}$, $\bar{p}^h_{\text{usv},e}$, the energy consumption can be reformulated by 
			\begin{equation}
				\begin{aligned}
					\chi_\Delta	&=
					\sum_{e=1}^{E+1}(m_e-n_{e-1})(\bar{p}_{\text{uav},e}+\bar{p}^f_{\text{usv},e})+\sum_{e=1}^{E} \\ & (n_e -m_e)  (p_h+\bar{p}^h_{\text{usv},e}).
				\end{aligned}
			\end{equation}
			
			In summary, let us collect the above constraints and denote variables by $\Theta_1 = \{E, m_e, n_e, r_k[m_e], \mathbf{q}[m_e],\mathbf{b}[m_e],\mathbf{b}[n_e] \}$, the hover point selection problem can be formulated by
			\begin{equation} \label{hov31}
				\begin{aligned}
					\mathrm{(P2):}&	
					\min_{ \Theta_1}\chi_\Delta \\
					\mathrm{s.t.}\quad
					&  \eqref{comconstraint}, \: n \in \mathcal{N}, \\
					&  \eqref{sensingconstraint}, \: \eqref{timeconstraint}, \: r_k[m_e]=\{0,1\}, \: \forall k, e. 
				\end{aligned}
			\end{equation}
			Solving problem $(\mathrm{P2})$ directly remains challenging due to the interplay between nonconvexity and discreteness. It manifests the following aspects: 1) The binary mixed-integer variable $r_{k}[m_e]$ introduces significant combinatorial complexity; 2) $r_{k}[m_e]$ is coupled with dwell time $n_e-m_e+1$ in \eqref{timeconstraint}, together determining the sensing strategy. 3) The decision dimension and feasible region vary with the unknown number of hover points $E$. This motivates us to develop a novel TSP-like approach in the next section. }

		\subsection{Three-step Hierarchical Hover Point Selection Method}
		{
			
			As shown in Fig. \ref{time2}, the solution procedure for $(\mathrm{P2})$ includes the following three steps.
			\begin{itemize}
				\item \textbf{Step 1}: The target schedule in each stage $e$ and rough hover point positions are determined by using a Virtual Base Station Coverage (VBSC) and clustering algorithm;
				\item \textbf{Step 2}: The visiting order sequence of the hover points is characterized by a BTSPN problem and solved by the open-loop TSP algorithm approximately; 
				\item \textbf{Step 3}: The durations of each stage and flying/hovering mode are optimized while refining the hover point positions by an SCA-based algorithm.  
			\end{itemize}
			Moreover, considering the UAV's hardware capability in practice, we make a further assumption about the sensing process.
			
			(A.IV) The number of targets that the UAV detects simultaneously in hovering mode is no more than $Z$.
			
			\textit{1) VBSC and Clustering Algorithm:} According to the maximum sensing targets assumption (A.IV), we can initially determine the number of hover points by
			\begin{equation}
				E_\text{in} = \lceil \frac{K}{Z} \rceil.
			\end{equation}
			Inspired by \cite{8255824}, we develop a VBSC and clustering algorithm with the given number $E_\text{in}$, whose pseudo-algorithm procedure is summarized in Algorithm \ref{alg:VBS_kmeans}. The core idea is to minimize the average distance between virtual base stations and their targets. Each virtual base station corresponds to one hover point, and under a coverage radius of $D_s$, each hover point can sense at most $Z$ targets. If the initial number $E_\text{in}$ does not meet the constraints, we will increase the number of hover points until the worse case $E_\text{in} = K$ considered in \cite{10680299}. Namely, the UAV has to hover above each target.    
			
			\begin{algorithm}[!t]
				\caption{VBSC and Clustering Algorithm}
				\label{alg:VBS_kmeans}
				\begin{algorithmic}[1]
					\State \textbf{Input:} Target set $\mathcal{K} = \{\mathbf{t}_1, \dots, \mathbf{t}_{K_\text{tar}}\}$, sensing distance $D_s$, max targets per cluster $Z$.
					\State \textbf{Output:} Final clusters $\{\mathcal{S}_i\}_{i=1}^{E}$ and centroids $\{\mathbf{c}_i\}_{i=1}^{E}$.
					\While{$E_\text{in} \leq {K_\text{tar}}$}
					\State Run $E_\text{in}$-means Algorithm to get $\{\mathcal{S}_i\}_{i=1}^{E_\text{in}}$ and $\{\mathbf{c}_i\}_{i=1}^{E_\text{in}}$.
					\State \textbf{Inner Loop 1:} \textbf{For} each cluster $i$
					\State \kern 10pt \textbf{If} {$|\mathcal{S}_i| > Z$} \textbf{then:} Redistribute the excess targets in $\mathcal{S}_i$ to other clusters and update centroids ${\mathbf{c}_i}_{i=1}^{E_\text{in}}$.
					\State \textbf{Inner Loop 2:} \textbf{For} each target $\mathbf{t}_j \in \mathcal{K}$
					\State \kern 10pt \textbf{Check }$\text{dist}(\mathbf{t}_j, \mathbf{c}_i) \leq D_s$ for all $\mathbf{t}_j \in \mathcal{S}_i$
					\State \kern 10pt \textbf{If yes:} Stop; \kern 5pt \textbf{If not:}  $ E_\text{in} \gets E_\text{in}+1$.
					\EndWhile
				\end{algorithmic}
			\end{algorithm}

			\textit{2) Bi-TSPN based Visiting Order Algorithm:} 
			Given the output of $E$ clusters and centroids in Algorithm \ref{alg:VBS_kmeans}, the visiting order sequence, which may significantly change the final trajectory, needs to be determined subsequently. This process resembles the traditional TSP with the neighborhood (TSPN) problem, but introduces a unique challenge by considering the collaboration of the UAV-USV. Specifically, different from the TSPN problem constructed for a single UAV in \cite{8255824}, two `salesmen' (UAV and USV) with different neighborhood radii ($D_s$ and $D_c$), velocities ($\mathbf{v}_\text{uav}$ and $\mathbf{v}_\text{usv}$) and cost indicators ($E_\text{uav}$ and $E_\text{usv}$) need to travel together and reach the end point simultaneously, which substantially complicates the solution. \textbf{We refer to this problem as a Bi-TSPN}, whose optimal solution, to our best knowledge, remains largely unexplored.        
			
			To approximately solve the Bi-TSPN problem, we construct a hybrid cost indicator that takes into account both the UAV's and USV's energy consumption. We assume that the UAV and USV move from one centroid $\mathbf{c}_i$ to another $\mathbf{c}_j$ at an average speed $\bar{v}_\text{uav}$ and $\bar{v}_\text{usv}$, and the velocity of the USV is given by $\mathbf{v}_{\text{usv}, i, j} = \bar{v}_\text{usv} (\mathbf{c}_j-\mathbf{c}_i)/d_{i,j}$, where $d_{i,j} = \sqrt{\|\mathbf{c}_j-\mathbf{c}_i\|^2}$ represents the distance. Recall the definitions of the power consumption in \eqref{fly11} and \eqref{h111}, the cost is formulated by \footnote{It should be noted that the power consumption of the UAV is dependent on its speed but is independent of its direction since the UAV flies in a free space. In contrast, the power consumption of the USV is affected by the angle between the direction of the USV's velocity and the direction of the water flow.}
			\begin{equation}
				E^\text{cost}_{i,j} = \frac{d_{i,j}}  {\bar{v}_\text{uav}}p^f_\text{uav}(\bar{v}_\text{uav})+\frac{d_{i,j}}{\bar{v}_\text{usv}} \alpha \|\mathbf{v}_{\text{usv},i,j} - \mathbf{v}_{w}(\mathbf{b})  \|^2,
			\end{equation}       
			for $i,j \in \{0,1,\cdots,E+1\}$. {Since water flow velocity is position-dependent, the water flow's influence on the USV can alternate between being a boost and a drag over the path from $\mathbf{c}_i$ to $\mathbf{c}_j$, as shown in Fig. \ref{shuiliu}. To characterize this phenomenon, we divided $d_{i,j}$ by the resolution $d_\text{wat}$ of the water current into $N_d = \lceil {d_{i,j}}/{d_\text{wat}} \rceil $ segments.} Thus, the energy cost can be recast by
			\begin{equation} \label{cost}
				E^\text{cost}_{i,j} = \frac{d_{i,j}}  {\bar{v}_\text{uav}}p^f_\text{uav}(\bar{v}_\text{uav}) +\sum_{k=1}^{N_d}\frac{d_{i,j}\alpha }{N_d\bar{v}_\text{usv}}  \|\mathbf{v}_\text{usv}- \mathbf{v}_{w}(\mathbf{b}_k)  \|^2,
			\end{equation}       
			where $\mathbf{b}_k$ is the start position of each segment. Once the cost between two arbitrary centroids is determined, we can apply the classic open-loop TSP algorithm to obtain the optimal sequence, as detailed in Appendix \ref{appA}.  
			
			\begin{figure}[!t]
				\centering
				\includegraphics[width=8cm]{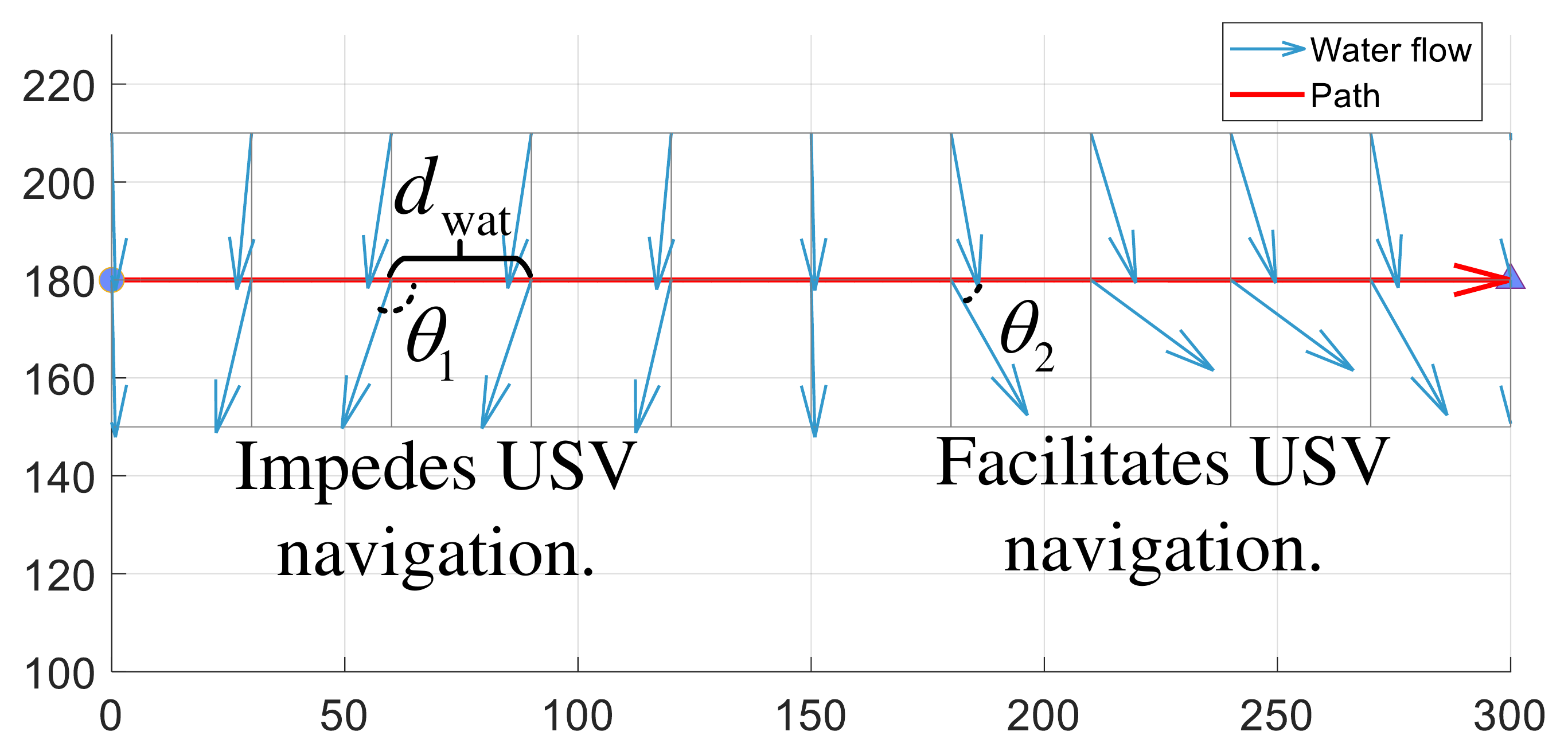}
				\caption{Variation of water flow direction along the path
					.}
				\label{shuiliu}
			\end{figure}
			
			\textbf{\textit{Remark 1}}: Unlike the traditional TSP which only considers the distance between two nodes, the cost in \eqref{cost} depends on the interaction of multiple factors such as the distance, the speeds and energy consumptions of UAV and USV, and the water current. It may form a completely distinguished visiting order when any of the factors change.

			\textit{3) Hover Point Refinement and Time Allocation Algorithm :} Note that the temporal correlations and the coordination of the UAV and USV are omitted in the previous procedures. In this step, we introduce the auxilary time variables $t^f_{e}$ (flying duration) and $t^h_{e}$ (hovering duration), and speed variables $\bar{v}_{\text{uav},e}$ (UAV average speed), $\bar{v}^f_{\text{usv},e}$ and $\bar{v}^h_{\text{usv},e}$ (USV average speed in flying and hovering mode) in the $e$-th stage. Based on the determined target schedule $r_{k,e}$ and the visiting order, we reformulate the original problem $(\mathrm{P1})$ to solve the auxiliary variables and refine the positions of hover points.
			
			First, to tackle the non-convex term of the UAV's energy consumption function with respect to (w.r.t) variable $\bar{v}_{\text{uav},e}$, we introduce an auxiliary variable  \(\xi_e \geq 0\) satisfying
			\begin{equation}\label{xi2}
				\xi^2_e=\sqrt{1+\frac{\bar{ v}_{\text{uav},e}^4}{4v_0^4}}-\frac{\bar{ v}_{\text{uav},e}^2}{2v_0^2}.
			\end{equation}
			Thus, the third term of \eqref{fly11} is recasted by $U_\text{uav}^1\xi_e$. Following the procedure in \cite{10680299}, equation \eqref{xi2} can be rewritten as  
			\begin{equation}
				\frac{1}{\xi^2_e}=\xi^2_e+\frac{\bar{ v}_{\text{uav},e}^2}{v_0^2}.
			\end{equation} 
			By using the first-order Taylor expansion, the right-hand side is replaced with its lower bound, as 
			\begin{equation}
				\begin{aligned}
					&	\frac{1}{\xi^2_e}  \leq
					(\xi^{(\kappa)}_e)^2+2\xi^{(\kappa)}_e(\xi_e-\xi^{(\kappa)}_e) +\frac{\bar{ v}_{\text{uav},e}^2}{v_{0}^{2}}
					\\&+
					\frac{2}{v_0^2}
					(\bar{{v}}^{(\kappa)}_{\text{uav},e})(\bar{{v}}_{\text{uav},e}-\bar{{v}}^{(\kappa)}_{\text{uav},e} )\triangleq g(\xi_e,\bar{{v}}_{\text{uav},e}),
					\label{hov7}
				\end{aligned}
			\end{equation}
			where the superscript $(\kappa)$ represents the iteration indicator.
			
			Second, by taking the UAV as an example, the positions of the hover points, the time and speed variables satisfies the equality $\| \mathbf q_e - \mathbf q_{e-1}\| =\bar{{v}}_{\text{uav},e} t^f_{e}$, which is a non-convex constraint. Again, we leverage the Taylor expression at the points $\bar{{v}}_{\text{uav},e}^{(\kappa)}$ and $ t_{e} ^{f(\kappa)}$, yielding
			\begin{equation}
				\begin{aligned}
					\|\mathbf{q}_e - \mathbf{q}_{e-1}\|
					&\leq {\bar{{v}}^{(\kappa)}_{\text{uav},e}} \, t^f_{e}
					+ {\bar{{v}}_{\text{uav},e}} \, {t_{e}^{f(\kappa)}}
					- {\bar{{v}}^{(\kappa)}_{\text{uav},e}} \, {t_{e}^{f(\kappa)}}.
				\end{aligned}
				\label{xin5}
			\end{equation}
			Similarly, the kinematic relationship for the USV can be expressed by
			\begin{equation}
				\begin{aligned}
					&\|\mathbf{b}_{e}^f - \mathbf{b}^h_{e-1}\|
					\leq {\bar{{v}}^{f(\kappa)}_{\text{usv},e}}\, t^f_{e}
					+ {\bar{{v}}^f_{\text{usv},e}} \, {t_{e}^{f(\kappa)}}
					- {\bar{{v}}^{f(\kappa)}_{\text{usv},e}} \, {t_{e}^{f(\kappa)}},\\
					&\|\mathbf{b}^h_{e} - \mathbf{b}^f_{e}\|
					\leq {\bar{{v}}^{h(\kappa)}_{\text{usv},e}} \, t^h_{e}
					+ {\bar{{v}}^h_{\text{usv},e}} \, {t_{e}^{h(\kappa)}}
					- {\bar{{v}}^{h(\kappa)}_{\text{usv},e}} \, {t_{e}^{h(\kappa)}},
				\end{aligned}
				\label{xin6}
			\end{equation}
			where $\mathbf{b}_{e}^f$ and $\mathbf{b}_{e}^h$ denote the endpoints of the flying mode and the hovering mode of the USV in the $e$-th stage, respectively.
			
			Third, for the other non-convex terms in the objective function, such as $t^f_{e} \bar{ v}_{\text{uav},e}^2$, we also conduct linearized operations at the points of $\kappa$-th iteration, for instance, 
			\begin{equation}
				\begin{aligned}
					\bar{{v}}_{\text{uav},e}^2t^f_{e} & =  (\bar{{v}}^{(\kappa)}_{\text{uav},e})^2 \, t_{e}^{f(\kappa)}
					+ (\bar{{v}}^{(\kappa)}_{\text{uav},e})^2
					(t^f_{e}-t_{e}^{f(\kappa)})\\
					&
					+2t_{e}^{f(\kappa)} \,\bar{{v}}^{(\kappa)}_{\text{uav},e}
					(\bar{{v}}_{\text{uav},e}-\bar{{v}}^{(\kappa)}_{\text{uav},e}).
				\end{aligned}
			\end{equation}
			The linearized objective function is denoted by $\chi_\Delta'$.
			
			In summary, denoting the optimization variables as $\Theta_2 = \{ t^f_{e}, t^h_{e}, \mathbf{q}_e,\mathbf{b}^f_{e},\mathbf{b}^h_{e},\xi_e,	\bar{{v}}_{\text{uav},e},	\bar{{v}}^f_{\text{usv},e},	\bar{{v}}^h_{\text{usv},e} \}$, the hover point refinement and time allocation problem is formulated by \footnote{In \eqref{timeconstraint}, the discrete time variable $n_e-m_e+1$ is replaced by $t_{h,e}$.}
			\begin{equation}
				\begin{aligned}
					\mathrm{(P3):}&
					\min_{  \Theta_2 }\chi_\Delta' \\
					\:\mathrm{s.t.} \quad & \eqref{sensingconstraint}, \: \eqref{timeconstraint},\:\eqref{xin5},\:\eqref{xin6}, \:  \\
					&\frac{1}{\xi^{2}_e} \leq g(\xi_e,\bar{{v}}_{\text{uav},e}),\\
					& \|\mathbf{q}_e-\mathbf{b}^f_{e}\| \leq D_c, \: \|\mathbf{q}_e-\mathbf{b}^h_{e}\| \leq D_c,  \\
					& \bar{{v}}_{\text{uav},e} \leq v_{\text{uav}}^{\text{max}}, \: \{\bar{{v}}_{\text{usv},h,e},\bar{{v}}_{\text{usv},f,e}\} \leq v_{\text{usv}}^{\text{max}},      
				\end{aligned}
			\end{equation}
			where $e=1,\cdots, E$. Problem $(\mathrm{P3})$ can be solved using CVX with the Mosek solver. Once the time variables $t^f_{e}$ and $t^h_{e}$ are obtained, the discretized time slots $n_e$ and $m_e$ can be determined by dividing the duration $\delta$.} 
		
		\section{Joint Trajectory Planning and Beamforming}	
		
		{The number, positions, and durations of the hover points have been determined in the previous section. The remaining work is to jointly plan the trajectories and design the beamforming matrices for UAV and USV in each stage. In this section, based on the similarity of each stage, we temporarily drop the subscript $e$ without causing ambiguity. For the sake of convenience, we denote $N_f$, $N_h$, and $K_h$ as the number of time slots for flying and hovering modes, and the number of targets assigned in each stage ($1 \le K_h \le Z$), respectively.}

		\subsection{Flying Mode Optimization}
		In flying mode, the UAV transmits a communication signal to the single-antenna USV without sensing interference. Observing the SNR expression in \eqref{ccc}, the optimal beamformer $\mathbf{w}_f[n]$ is indeed the MRT-based solution in assumption (A.I). By substituting the MRT beamformer, the start and end points of $e$-th stage, the target schedule, and the flying and hovering duration into problem $(\mathrm{P1})$, the to-be-optimized variables reduce to the trajectories $\mathbf{q}[n]$,$\mathbf{b}[n]$ and power allocation $p_c[n]$, whose optimization problem can be formulated by
		\begin{equation}
			\begin{aligned}
				\mathrm{(P4):}&	\min_{\{
					\mathbf{q}[n], \mathbf{b}[n],   p_c[n]
					\}  }  E^{\text{total}}_{\text{uav}}+E^{\text{total}}_{\text{usv}} \\
				\mathrm{s.t.} \kern 10pt  &  \eqref{p0-3}, \: \eqref{p18h} \:, n =1,2,\cdots, N_f,
				\\
				&	0\leq  p_c[n] \leq p_{\text{max}}, \: n =1,2,\cdots, N_f.
			\end{aligned}
		\end{equation}
		Here, the remaining constraints are the obstacle avoidance, the speed limitation, and the communication rate requirement. Problem $(\mathrm{P4})$ is convex and hence can be solved by the off-the-shelf CVX toolbox.
		
		\subsection{Hovering Mode Optimization}
		In hovering mode, however, the MRT beamformer is not applicable due to the presence of sensing interference. Moreover, only the trajectory of the USV needs to be determined since the position of the UAV is fixed at the hover point. Consequently, the optimization problem for joint trajectory and beamforming design can be formulated by
		\begin{equation} \label{hoverJoint}
			\begin{aligned}
				\mathrm{(P5):}	&\min_{\{
					\mathbf{b}[n], \mathbf{w}_h[n], \mathbf{v}_k[n]
					\}  }  E^{\text{total}}_{\text{uav}}+E^{\text{total}}_{\text{usv}} \\
				\mathrm{s.t.} \kern 10pt  &  \eqref{p0-3}, \: \eqref{p18h},
				\: {R}_h[n] \geq  \Gamma_h, \\
				&	\sum_{n=1}^{N_h}[n]\gamma_k[n] \geq  \Gamma _s^{\text{total}},  \: k = 1,..,K_h,
				\\ & \|\mathbf{w}_h[n]\|^2 +	\sum_{k=1}^{K_h} \|\mathbf{v}_k[n]\|^2\leq p_{\text{max}}.
			\end{aligned}
		\end{equation}
		In what follows, we propose an alternating optimization algorithm that iteratively optimizes the trajectory $\mathbf{b}[n]$ and the beamforming vectors $ \{\mathbf{w}_h[n], \mathbf{v}_k[n]\} $. 
		
		\subsubsection{Beamforming Design} 
		For a fixed trajectory $\mathbf{b}^{(\kappa)}[n]$ in the $\kappa$-th iteration, we introduce the auxiliary rank-$1$ positive semi-definite matrices $\mathbf{W}_h[n]=\mathbf{w}_{{h}}[n]\mathbf{w}^H_{{h}}[n]$ and $ \mathbf{V}_{{k}}[n]=r_k[n]\mathbf{v}_{{k}}[n]\mathbf{v}^H_{{k}}[n] $ and transform the original problem into an semi-definite programming (SDP) problem. By denoting the effective S\&C channels as $ \tilde{\mathbf{H}}_{k}[n] = \mathbf{H}^H_{k}[n] {{{\mathbf{u}}}}_k[n] {{{\mathbf{u}}}}_k^H[n] \mathbf{H}_{k}[n] $ and $\mathbf{H}_{c}[n]=\mathbf{h}_{c}[n]\mathbf{h}^H_{c}[n]$, respectively, the beamforming problem can be constructed by
		\begin{subequations} \label{Beamhover}
			\begin{align}
				\mathrm{(P6):}&	\min_{ \{\mathbf{W}_h[n],\mathbf{V}_{k}[n] \} }  \kern 5pt  \sum_{n=1}^{N_h} \left( \mathrm{tr} (\mathbf{W}_h[n]) + \sum_{k=1}^{K_h} \mathrm{tr}(\mathbf{V}_{k}[n]) \right) \notag \\
				\mathrm{s.t.}\quad &{R}_h[n] \geq \Gamma_h,  \:  \sum_{n=1}^{N_h}\gamma_k[n] \geq \Gamma_s^{\text{total}},  \: k = {1,...,K_h},	\\
				& \sum_{k=1}^{K_h}\mathrm{tr}(\mathbf{V}_{{k}}[n])+\mathrm{tr}(\mathbf{W}_h[n])\leq p_{\text{max}}, \\
				&\mathbf{W}_h[n] \succeq \mathbf{0},\: \mathbf{V}_k[n] \succeq \mathbf{0}.
			\end{align}
		\end{subequations}
		With a given trajectory, the problem $(\mathrm{P5})$ reduces to minimizing the beamforming energy consumption while meeting the S\&C performance requirements. Problem $(\mathrm{P6})$ forms a standard convex beamforming design in the ISAC field, which can be solved by the classical SDR method \cite{5447068}. The beamforming vector can be immediately reconstructed by an approach such as eigenvalue decomposition.  
		
		\subsubsection{USV Trajectory Design} For the fixed beamforming design $\{
		\mathbf{w}^{(\kappa)}_h[n],\mathbf{v}^{(\kappa)}_{k}[n] 
		\}$,  the USV trajectory design can be formulated by  
		\begin{subequations} \label{rty}
			\begin{align}
				\mathrm{(P7):}&	\min_{\{\mathbf{b}[n] \}} \kern 5pt   \sum_{n=1}^{N_h} \|\mathbf{v}_\text{usv}[n]- \mathbf{v}_{w}[n]   \|^2 \notag\\
				\mathrm{s.t.} \kern 5pt  & \eqref{p0-3}, \: \eqref{p18h},\:	{R}_h[n] \geq \Gamma_h. \notag 
			\end{align}
		\end{subequations}
		The UAV energy consumption and the sensing constraint disappear since they are unrelated to the motion trajectory of the USV. 
		The non-convexity of problem $(\mathrm{P7})$ comes from the communication constraint, where the communication channel $\mathbf{h}_c[n]$ depends on the positions of UAV (fixed) and USV. After simple algebraic operations, the communication constraint is equivalently transformed to 
		\begin{equation} \label{hovercomcst}
			\|\mathbf{h}^H_c[n]\mathbf{w}_f[n]\|^2 - \frac{\varrho N_s t_p}{\delta}\sum_{k=1}^{K_h} \|\mathbf{h}^H_c[n]\mathbf{v}_k[n]\|^2 \geq \varrho \sigma_h^2,
		\end{equation}
		where $\varrho=\frac{\delta}{N_st_p}(2^{{ \Gamma _h }{}}-1)$. Let us denote the function of the left-hand side as $f(\mathbf{b}[n])$. Similarly, by applying the SCA approach again, constraint \eqref{hovercomcst} is approximately transformed to a convex one, i.e.,
		\begin{equation}
			f(\mathbf{b}^{(\kappa)}[n])+\nabla f(\mathbf{b}^{(\kappa)}[n])^T(\mathbf{b}[n]-\mathbf{b}^{(\kappa)}[n]) \geq \varrho\sigma_h^2.
			\label{p3-11}
		\end{equation}
		{
			Again, problem $(\mathrm{P7})$ can be efficiently solved by the CVX toolbox. The procedure of the proposed alternating optimization algorithm is summarized in Algorithm \ref{AO}. 
		}

		\begin{algorithm}[t]
			\caption{Alternating Optimization Algorithm}
			\begin{algorithmic}[1]
				\State \textbf{Input:} 
				$\mathbf{t}_k$, $\tilde{\mathbf{t}}_{k'}$, $r_k[n]$, $\mathbf{q}[n]$, $\Gamma_h$, $\Gamma^{\text{total}}_s$, $\varepsilon=10^{-3}$.
				\State \textbf{Output:} $\mathbf{b}[n]$, $\mathbf{w}_h[n]$, $\mathbf{v}_{k}[n]$, $n = 1,\dots,N_h$.
				\Repeat
				\State Solve subproblem $\mathrm{P6}$ for the given  ${\mathbf{b}^{(\kappa-1)}[n]}$, and obtain 
				$\{{\mathbf{w}_h}^{(\kappa)}[n], {\mathbf{v}_k}^{(\kappa)}[n]\}$;
				\State Solve subproblem $\mathrm{P7}$ for the given $\{{\mathbf{w}_h}^{(\kappa)}[n], {\mathbf{v}_k}^{(\kappa)}[n]\}$, and obtain ${\mathbf{b}^{(\kappa)}[n]}$;
				\State Set $\kappa = \kappa+1$;
				\Until{the objective value converges within the threshold $\varepsilon$ or the maximum number of iterations $T_{\text{max}}$ is reached.}
			\end{algorithmic}
			\label{AO}
		\end{algorithm}

		\section{Numerical Simulation}
		\begin{table}[]
			\centering
			\caption{System parameters }
			\begin{tabular}{|c|c|c|c|}
				\hline
				\textbf{Parameter} & \textbf{Value} & \textbf{Parameter} & \textbf{Value} \\ \hline
				\makecell{$K$} & $15$ &\makecell{$H$} & $100 $ m	  \\ \hline
				\makecell{$M$} & $4$ &\makecell{$T_s$} & $0.01 $ s	  \\ \hline
				\makecell{$t_p$} & $0.005$ s&\makecell{$t_o$} & $0.005$ s	\\ \hline
				\makecell{$\delta$} & $1$ s &\makecell{$N_s$} & $100$\\ \hline
				\makecell{$\rho_0$} & $14.8$ dBm 
				&\makecell{$\sigma_c^2$} & $-110$ dBm\\ \hline	
				\makecell{$\beta $} & $14.8\:$dBm &\makecell{$\sigma_s^2$,$\sigma_h^2$ } & $-110$ dBm\\ \hline

				\makecell{$v_{\text{usv}}^{\text{max}}$} & $10$ m/s &\makecell{$v_{\text{uav}}^{\text{max}}$} & $20 $ m/s\\ \hline	
				\makecell{$p_s $} & $5$ W&\makecell{$p_c $} & $5$ W\\ \hline	
				
				\makecell{$p_{\text{max}}$} & $20$ W &\makecell{ $\alpha$  } & $20 $ kg\\ \hline
				\makecell{$\eta $} & $0.1 \: \text{m}^2$ &\makecell{$\epsilon$} &$0.001$ \\ \hline	
				
				\makecell{$U_\text{uav}^0$} & $80\:$ W &\makecell{$U_{\text{tip}}$} & $120 \:$	  rad/s \\ \hline
				\makecell{$d_0$} & $0.6\:$ rad&\makecell{$\rho$} & $1.225 \: $kg/$\text{m}^3$ 	\\ \hline
				\makecell{$\varphi$} & $0.05\:\: \text{m}^3$  &\makecell{$A$} & $0.503\:\:\text{m}^2$\\ \hline
				\makecell{$\Gamma _s^{\text{total}}$} & $12\:$dB  &\makecell{$\Gamma _s$} & $3 \:$dB \\ \hline
				\makecell{$\Gamma _c$} & $13\:$bps/HZ &\makecell{$Z$} &$8$ \\ \hline
			\end{tabular}
			\label{canshu}
		\end{table}
		

		{
			In this section, we show the effectiveness of the proposed air-sea collaborative framework through numerical simulations. The simulation parameters are listed in Table \ref{canshu}, and we set $\Gamma_c = \Gamma_h = \Gamma_f$ for simplicity. The baseline approaches are as follows.
			
			\begin{itemize}[]
				\item
				\textit{Sequential Access Strategy}: The strategy in \cite{10680299} adopts a similar flying and hovering mode for a single UAV ISAC system, where the UAV must hover over the targets. This scenario corresponds to the worst case of our scheme, where the sensing distance is set by $D_S = H$.    
				\item
				\textit{Leader–follower Strategy}: As this is the first work for the ISAC-enabled UAV-USV collaborative system, we choose a leader-follower strategy to show the performance gain of the joint design scheme. Namely, the trajectory of the UAV is first optimized by using the TSPN method in \cite{8255824}, followed by the optimization of the USV's trajectory.     	
			\end{itemize}
			To guarantee the fairness of the comparison, we have added and completed the other system designs that the baseline methods omitted, such as beamforming, etc. 
			
			
			\subsection{The Superiority of Energy Efficiency}

			\begin{figure*} [t]
				\centering
				\subfloat[\normalfont \label{newduibi1}Our scheme]{
					\includegraphics[width=4.5cm]{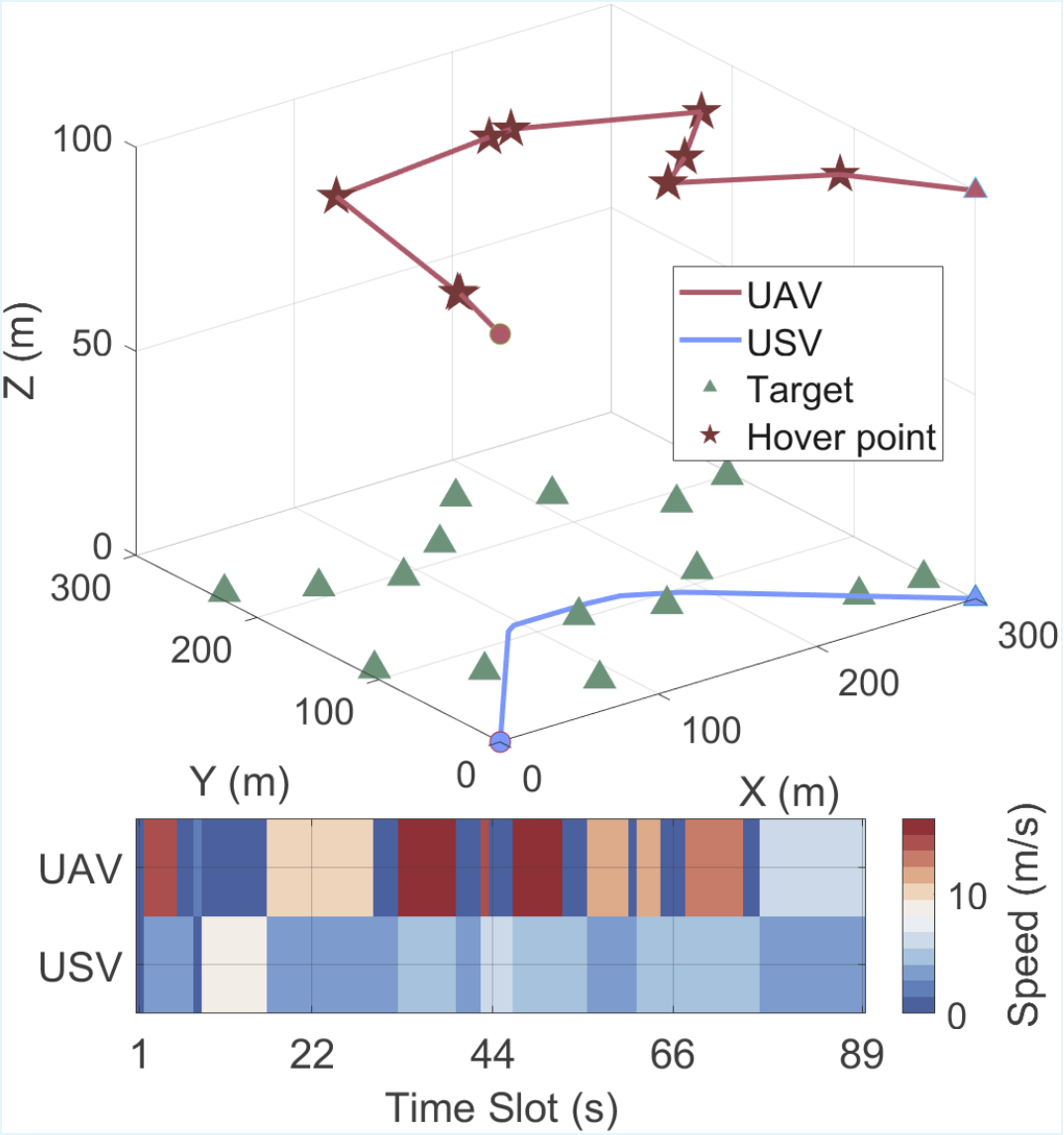}}
				\subfloat[\normalfont\label{newduibi2} Sequential Access]{
					\includegraphics[width=4.5cm]{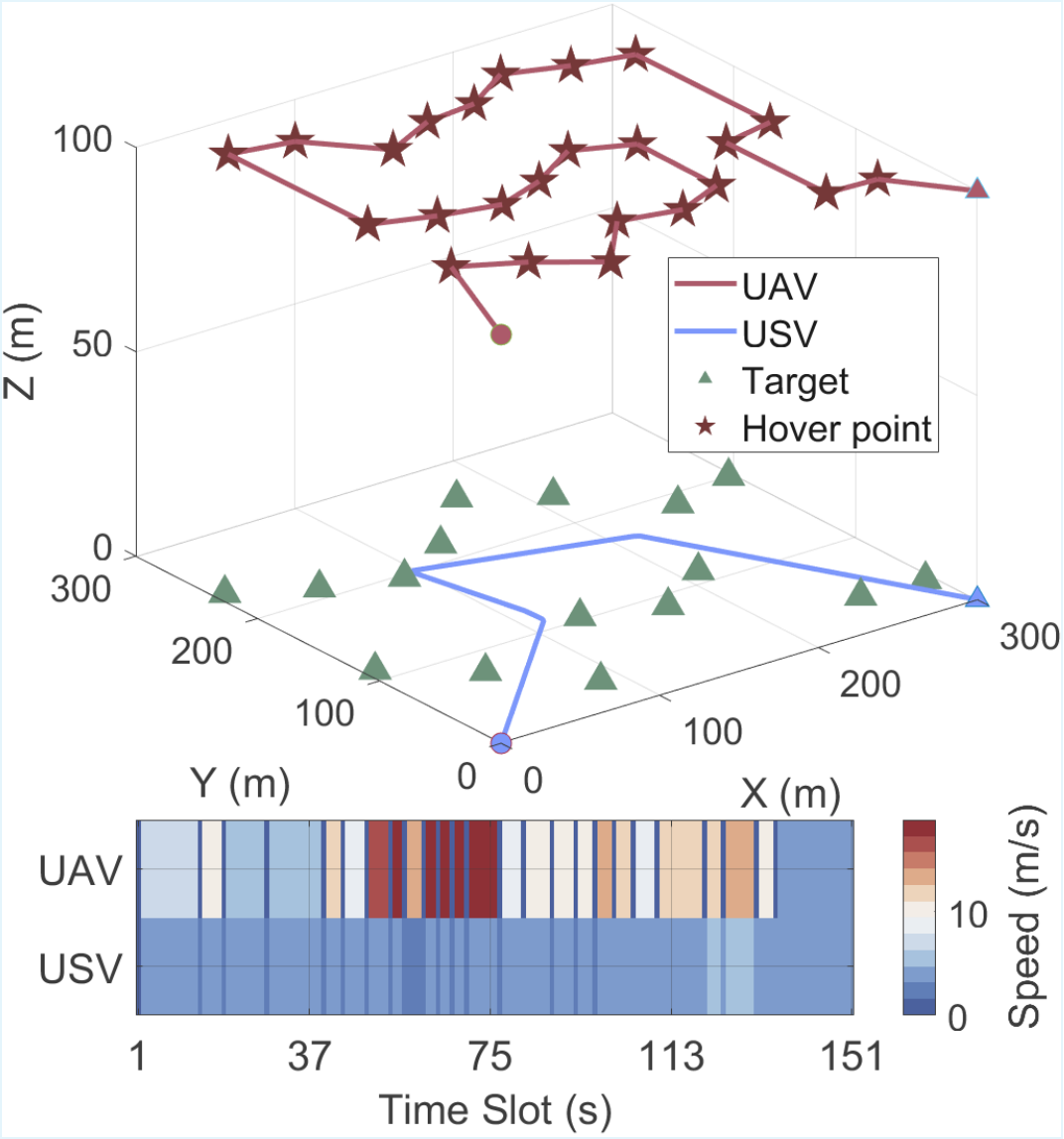}}
				\subfloat[\normalfont\label{newduibi3} Leader–Follower]{
					\includegraphics[width=4.5cm]{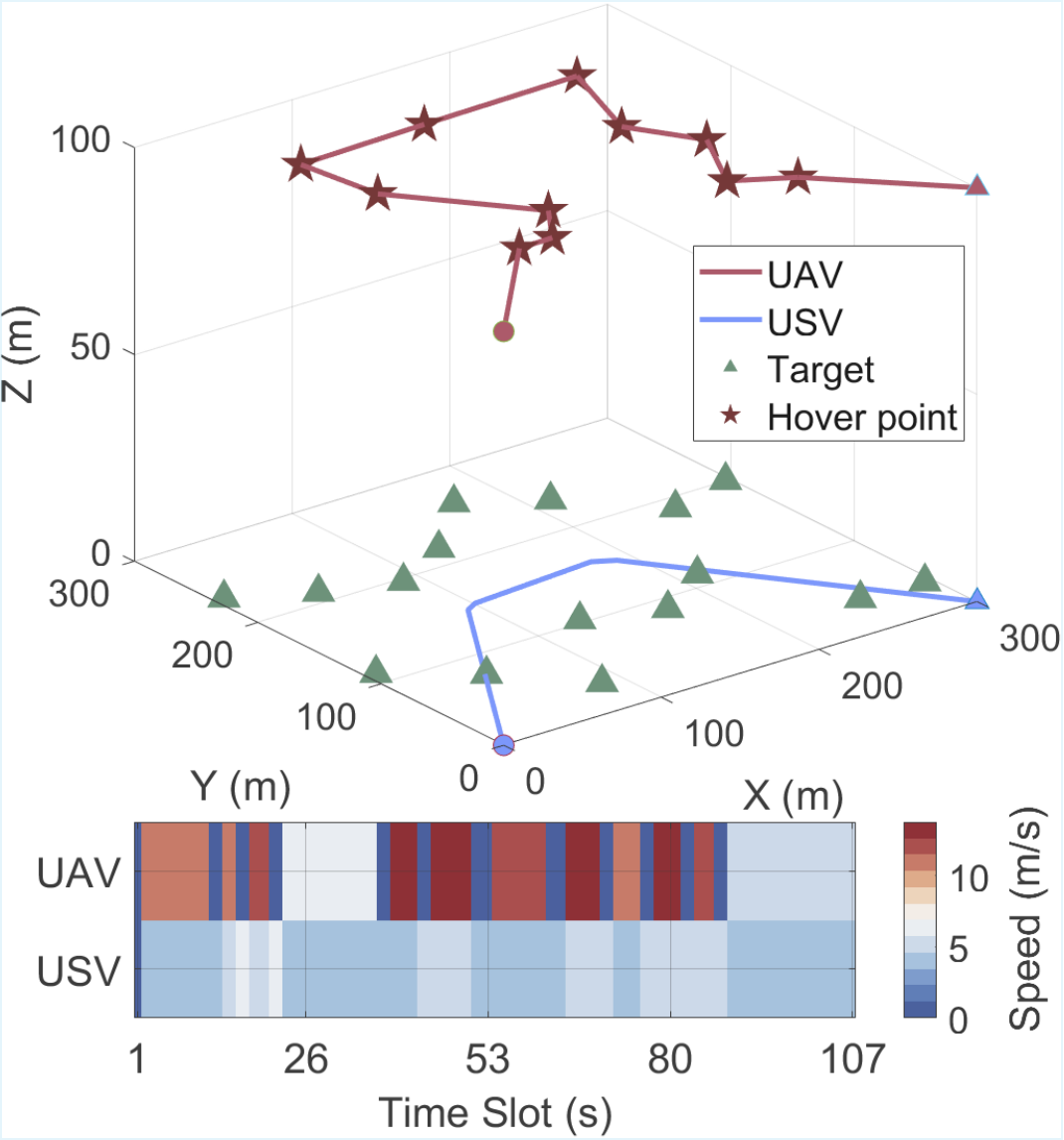}}
				\subfloat[\normalfont\label{newduibi4} Top view of our scheme]{
					\includegraphics[width=4.6cm]{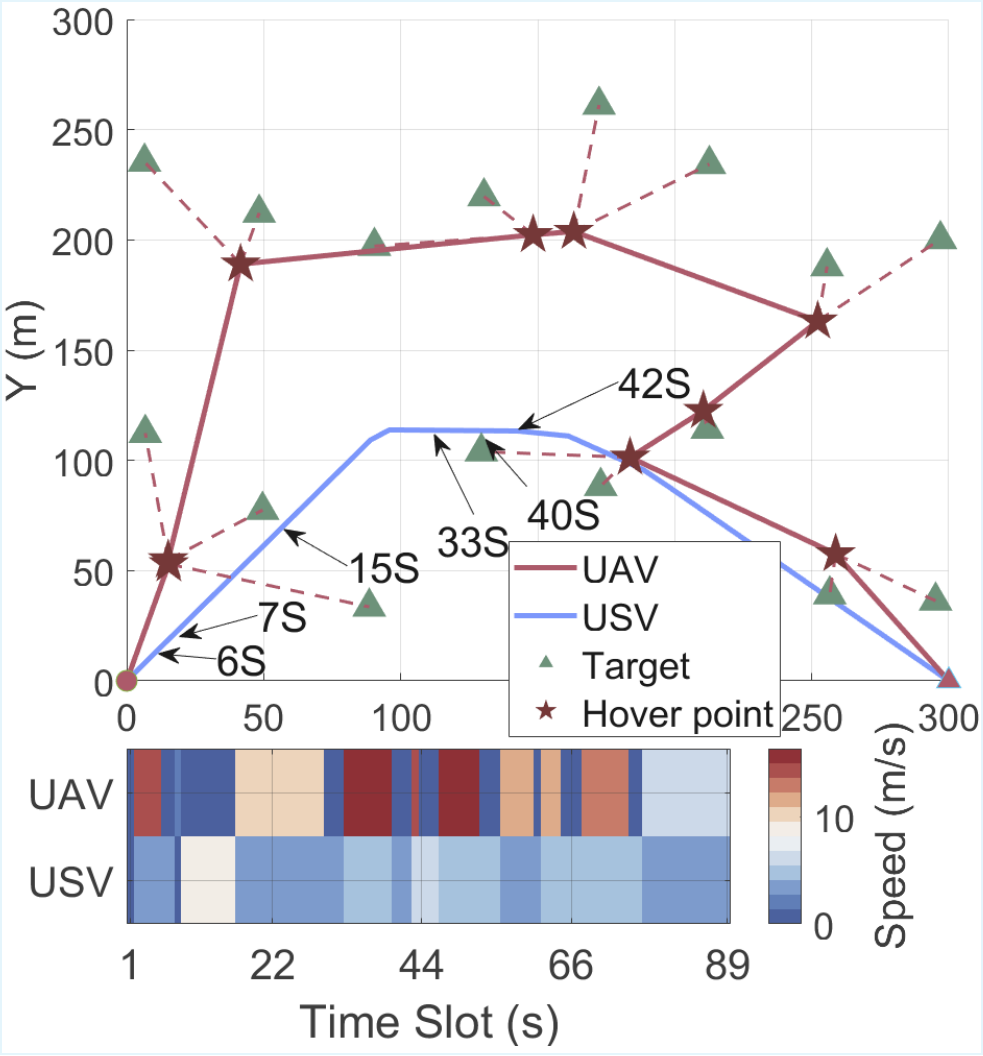}}
				
				\caption{The comparisons for the trajectories of UAV and USV.
				}
				\label{We} 
			\end{figure*}
			
			Figs.~\eqref{newduibi1},~\eqref{newduibi2}, and \eqref{newduibi3} show the trajectories of the proposed scheme, sequential access strategy, and leader–follower strategy, respectively. The UAV and USV start at $(0, 0)$ m and end at $(300, 300)$ m in the horizontal direction with $H=100$m. The dashed lines indicate the matching of hover points with sensing targets. The total energy consumption for the three schemes is $40.91$ kJ (our scheme), $58.06$ kJ (sequential access), and $50.94$ kJ (leader-follower), respectively. It shows the superiority of our scheme in terms of energy efficiency. The reason behind this is that the UAV visits many fewer hover points and travels a much shorter path than the sequential access strategy. Moreover, the trajectory planning of the UAV does not consider the influence of the USV, which must follow a fixed UAV path to guarantee the communication requirement, resulting in wasted energy consumption by the USV. In contrast, the UAV and USV travel synchronously and coordinate with each other in our scheme.     
			
			
		}

		\begin{figure} [t!]
			\centering
			\subfloat[\normalfont\label{SCa}Power varies over time with $p_{\text{max}} = 20$ W.
			]{
				\includegraphics[width=8.8cm]{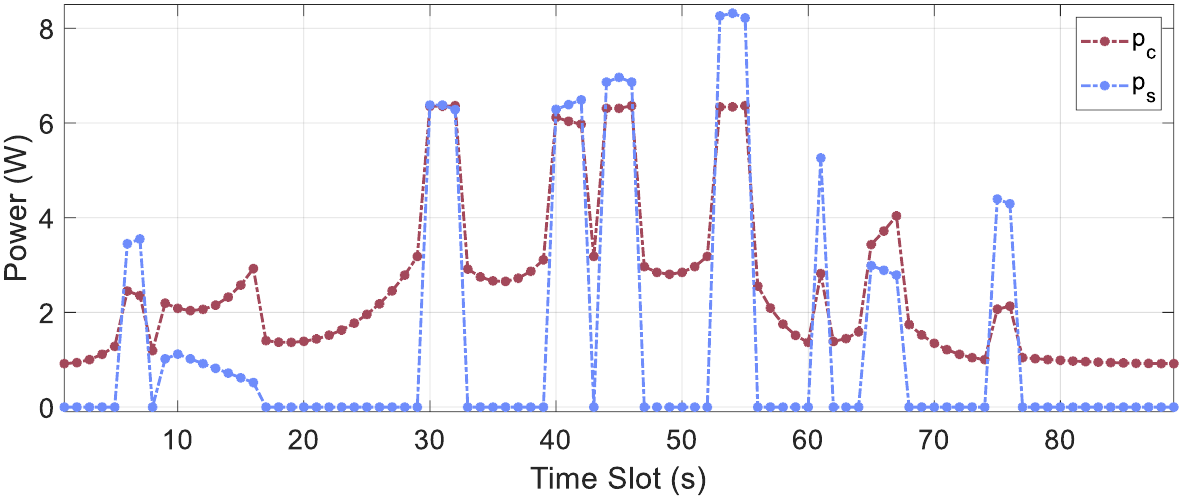}}
				
			\subfloat[\normalfont\label{SCc}Power varies over time with $p_{\text{max}} = 15$ W.
			]{\includegraphics[width=8.8cm]{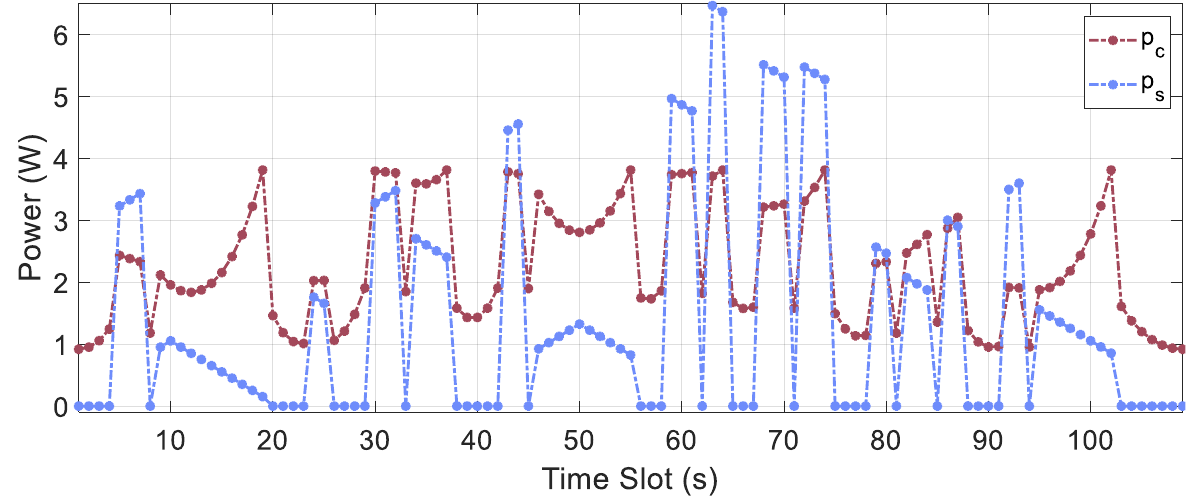}}
			
			\subfloat[\normalfont\label{SCb}
Top view in the $4$-th stage with $p_{\text{max}} = 20$ W.
			]{\includegraphics[width=8.9cm]{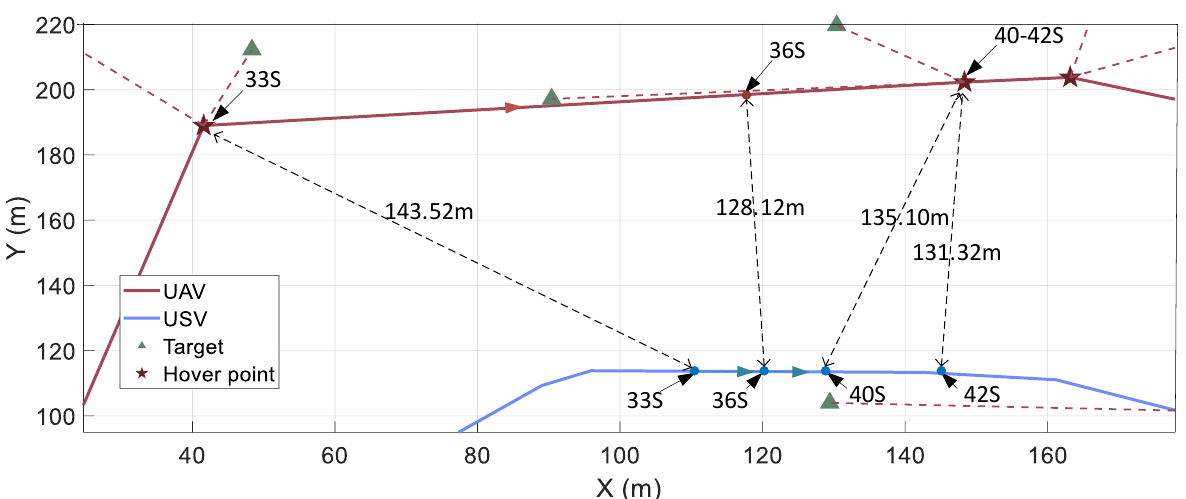}}
			\caption{The changes in S\&C power.}
		\end{figure}
		
		\subsection{The Power Allocation for S\&C Functions }
		
Fig. (\ref{SCa}) shows the variation of S\&C power over time. 
For example, in the $4$-th stage ($33-42$s), the actual positions of the UAV and USV are shown in Fig. (\ref{SCb}).
From $33$s to $36$s, the UAV and USV move along the arrow direction, with the distance decreasing from $143$ m to $128$ m, and the communication power decreasing accordingly. From $36$s to $40$s, as the distance increases, the communication power also increases. From $40$s to $42$s, the system enters hovering mode, with the UAV's position remaining unchanged while the USV moves closer to the UAV, causing the communication power to decrease.
In different hover modes, the sensing power depends on hover time, the number of sensing targets, and the distance to the hover point. 
Furthermore, the variation in sensing power is related to communication power, and they have a competitive relationship. Since the total power is fixed, higher communication power results in lower remaining sensing power.

To assess the effect of the power limit, we set it to 15 W, as shown in Fig. (\ref{SCc}).
As the power limit decreases, both hover time and the number of hover points increase. 
This is because, with a lower power limit, sensing powers decrease, and to meet the instantaneous SNR requirements, hover points must be placed closer to sensing targets, requiring more time to accumulate the SNR. 

		\subsection{Impact of the Number and Distribution of Targets}

		\begin{figure} [t!]
			\centering
			
			\subfloat[\normalfont\label{We2}$K = 32$]{
				\includegraphics[width=4.5cm]{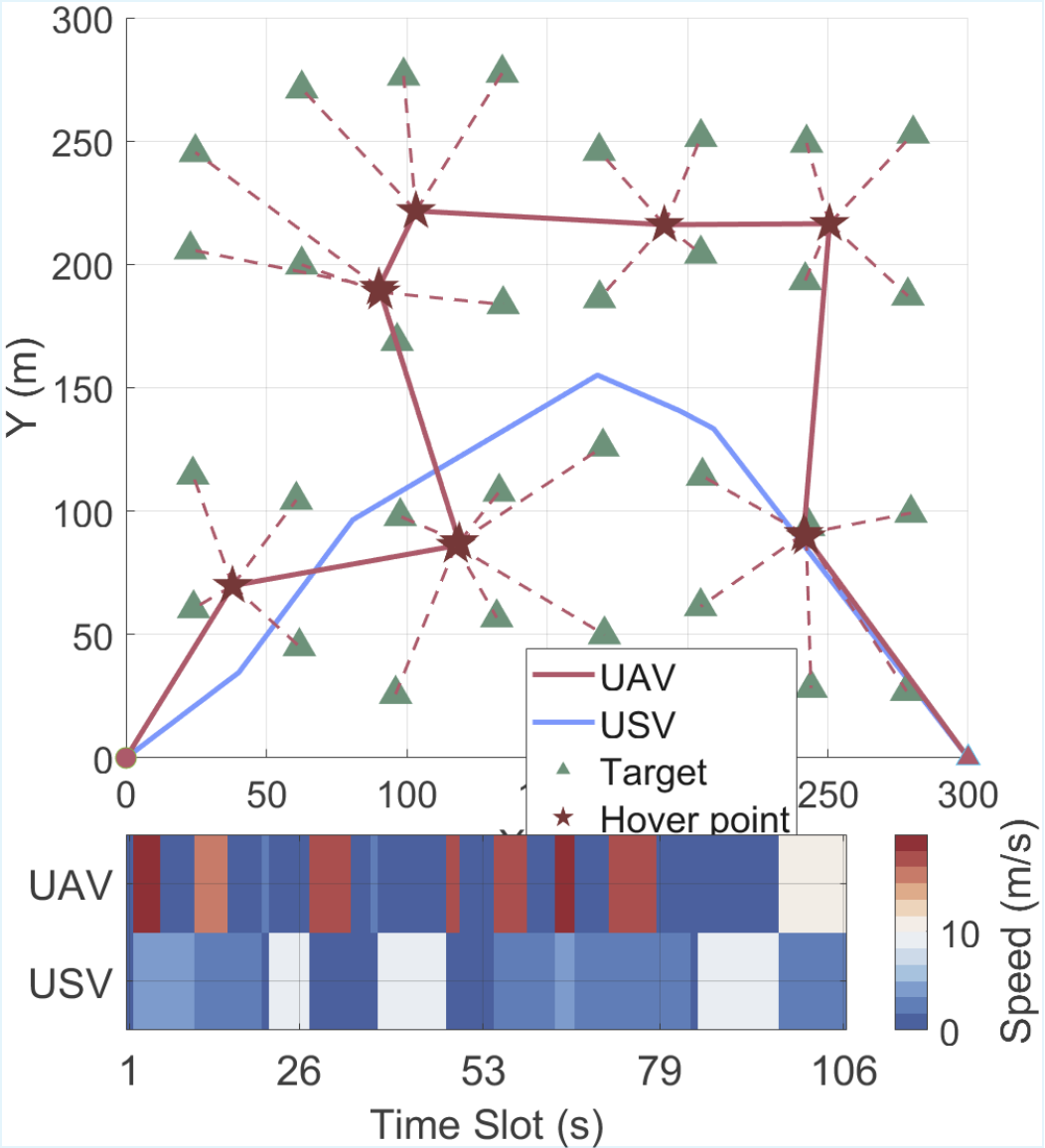}}
			\subfloat[\normalfont\label{We1} $K = 56$]{
				\includegraphics[width=4.5cm]{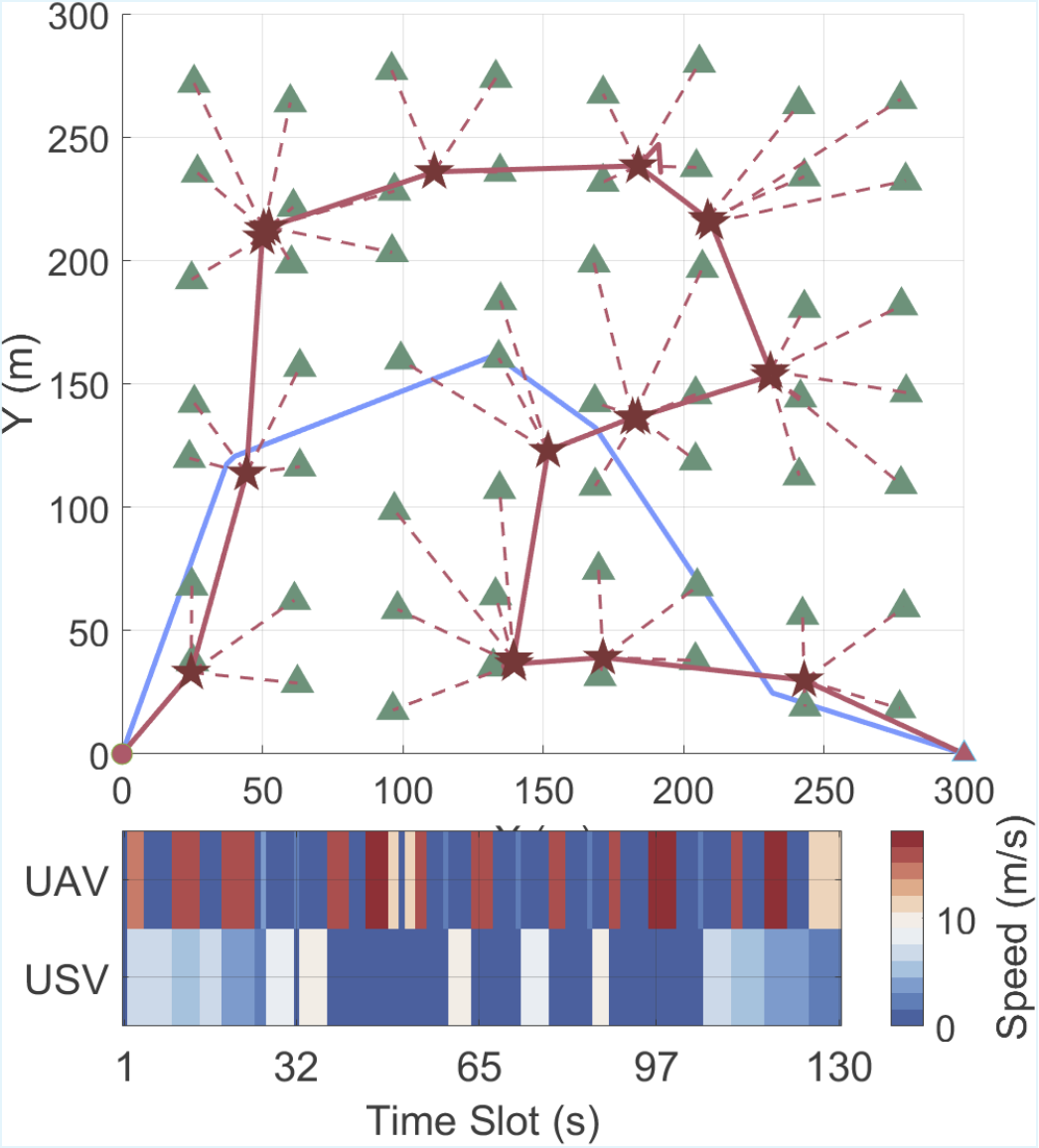}}		
			\caption{Top view of the trajectory under different $K$ values. 
			}
			\label{WeK_sum} 
		\end{figure}

					\begin{figure} [t!]
			\centering
			
			\subfloat[\normalfont\label{We2}$\sigma = 30\: $m]{
				\includegraphics[width=4.5cm]{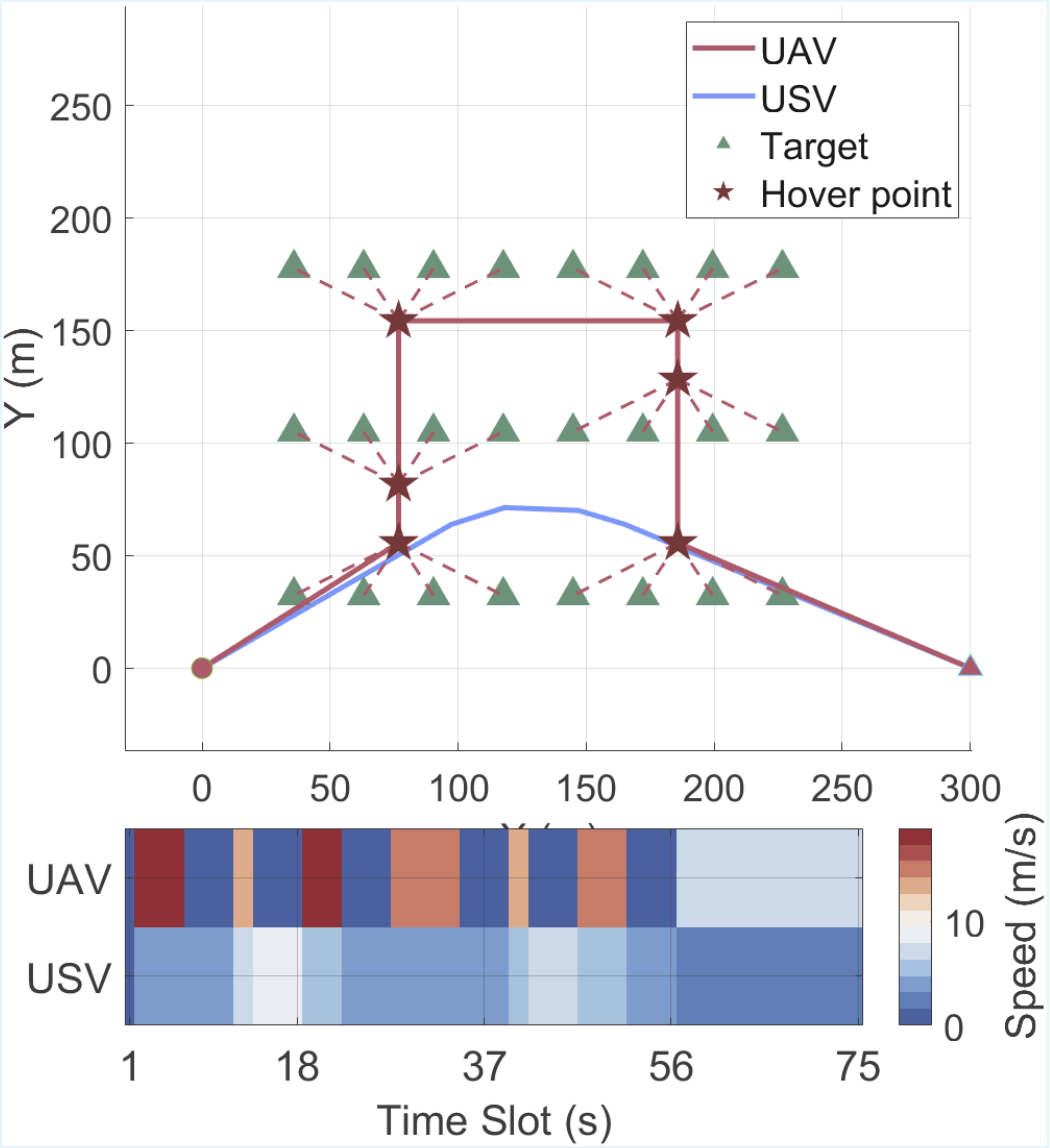}}
			\subfloat[\normalfont\label{We1} $\sigma = 50 \:$m]{
				\includegraphics[width=4.5cm]{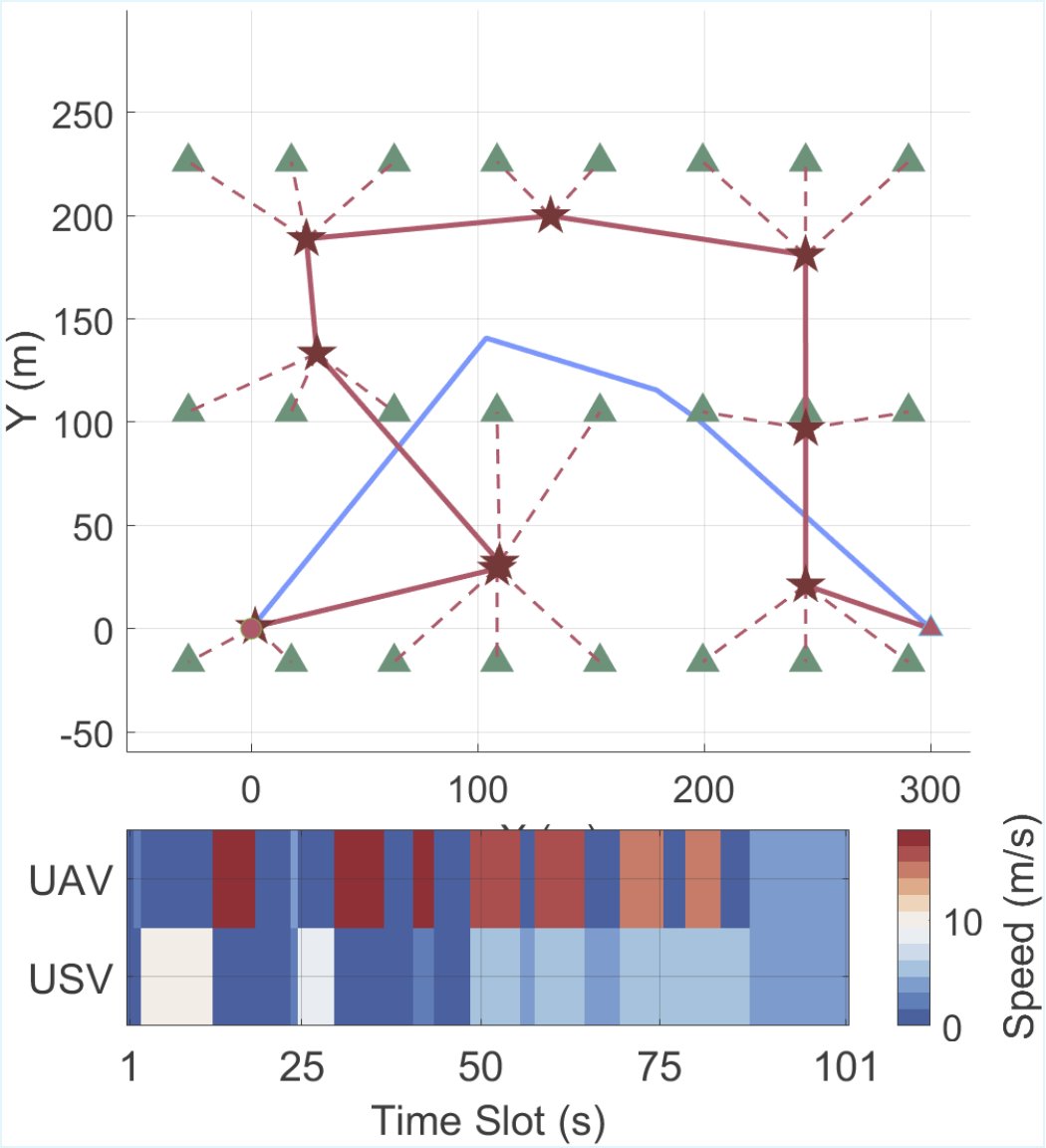}}		
			\caption{Top view of the trajectory with different $\sigma$. 
			}
			\label{WeK_fangcha} 
		\end{figure}

		\begin{figure} [t!]
			\centering
			\subfloat[\normalfont \label{We1} $\Gamma _{s} = 1\:$dB]{
				\includegraphics[width=4.5cm]{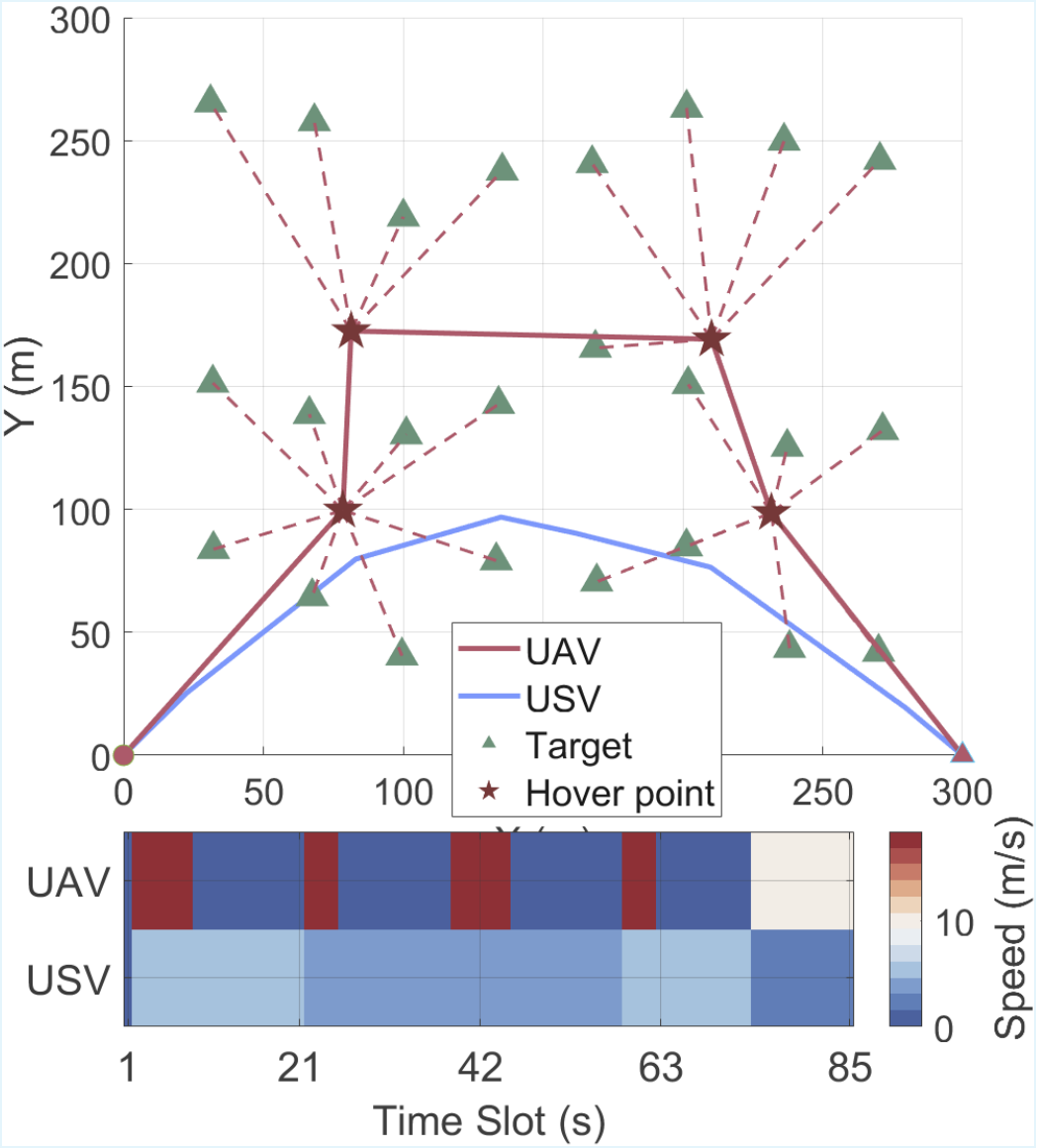}}
			\subfloat[\normalfont \label{We1} $\Gamma _{s} = 5\:$dB]{
				\includegraphics[width=4.5cm]{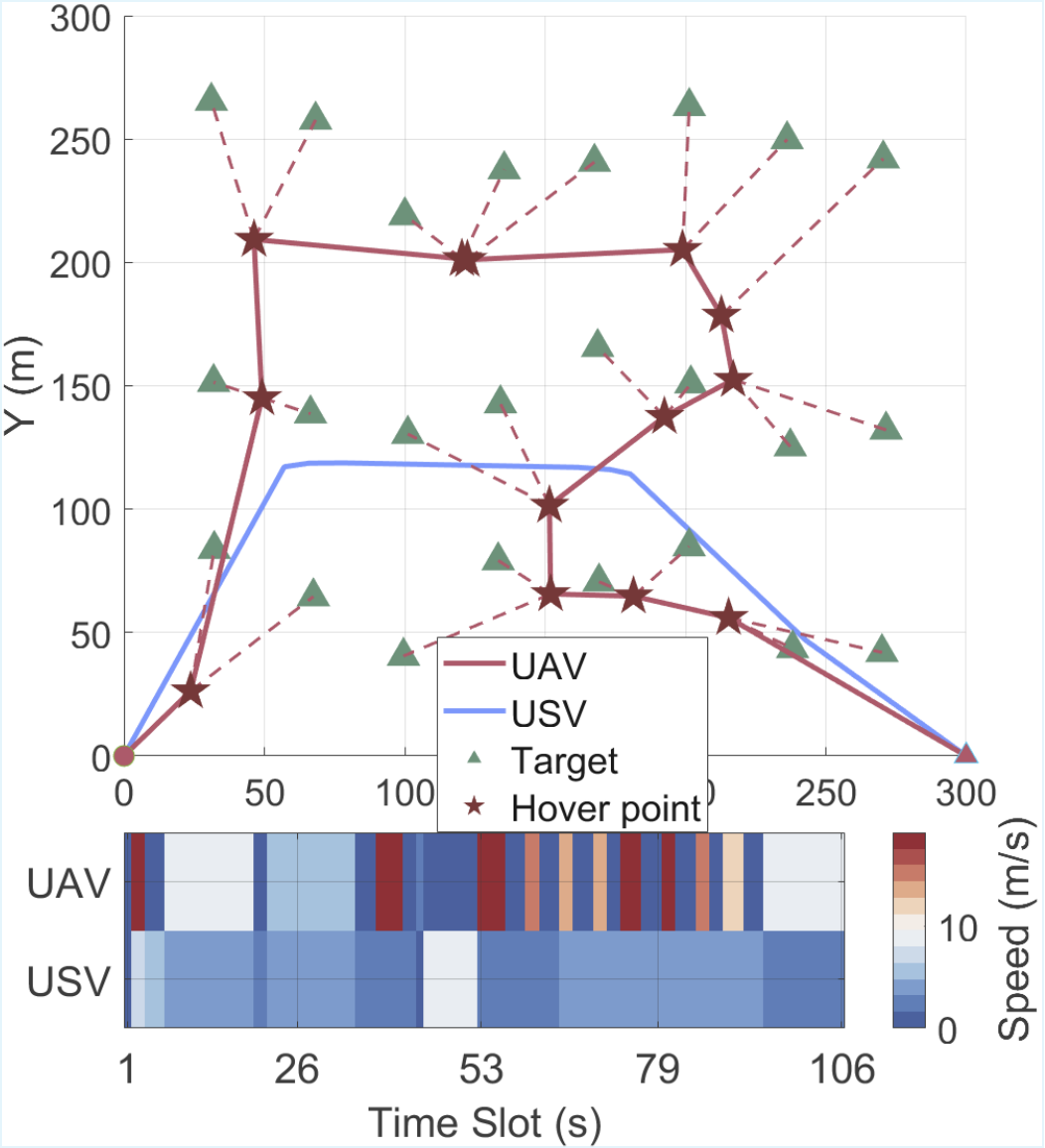}}			
			\caption{The system trajectory at different $\Gamma _{s}$.
			}
			\label{WeS} 
		\end{figure}
		
		\begin{figure} [t!]
			\centering
			\subfloat[\normalfont \label{w3} $		v_{\text{max}}^w = 3\:$ m/s]{
				\includegraphics[width=4.5cm]{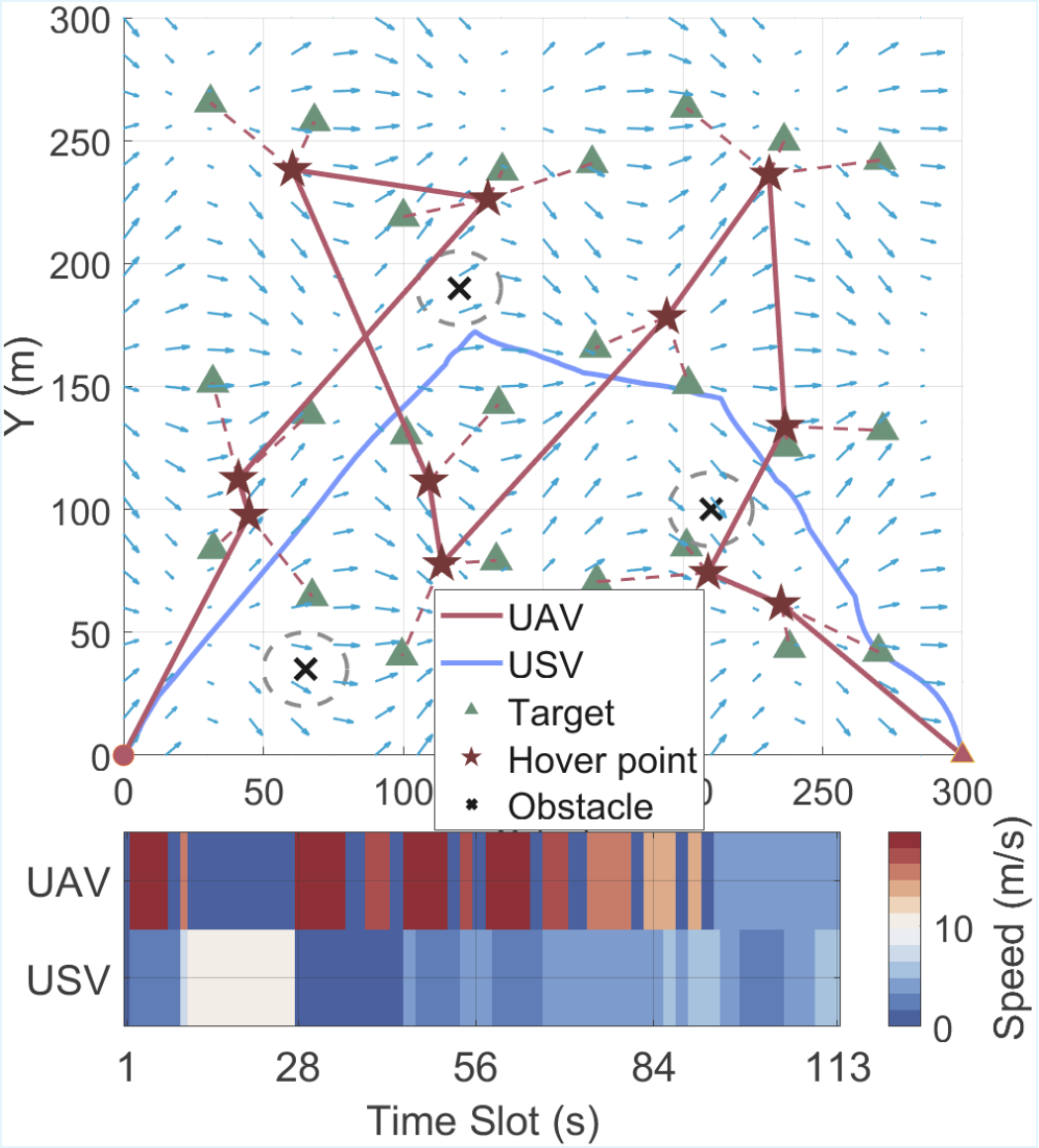}}
			\subfloat[\normalfont \label{w-3} $		v_{\text{max}}^w = -3\:$ m/s]{
				\includegraphics[width=4.5cm]{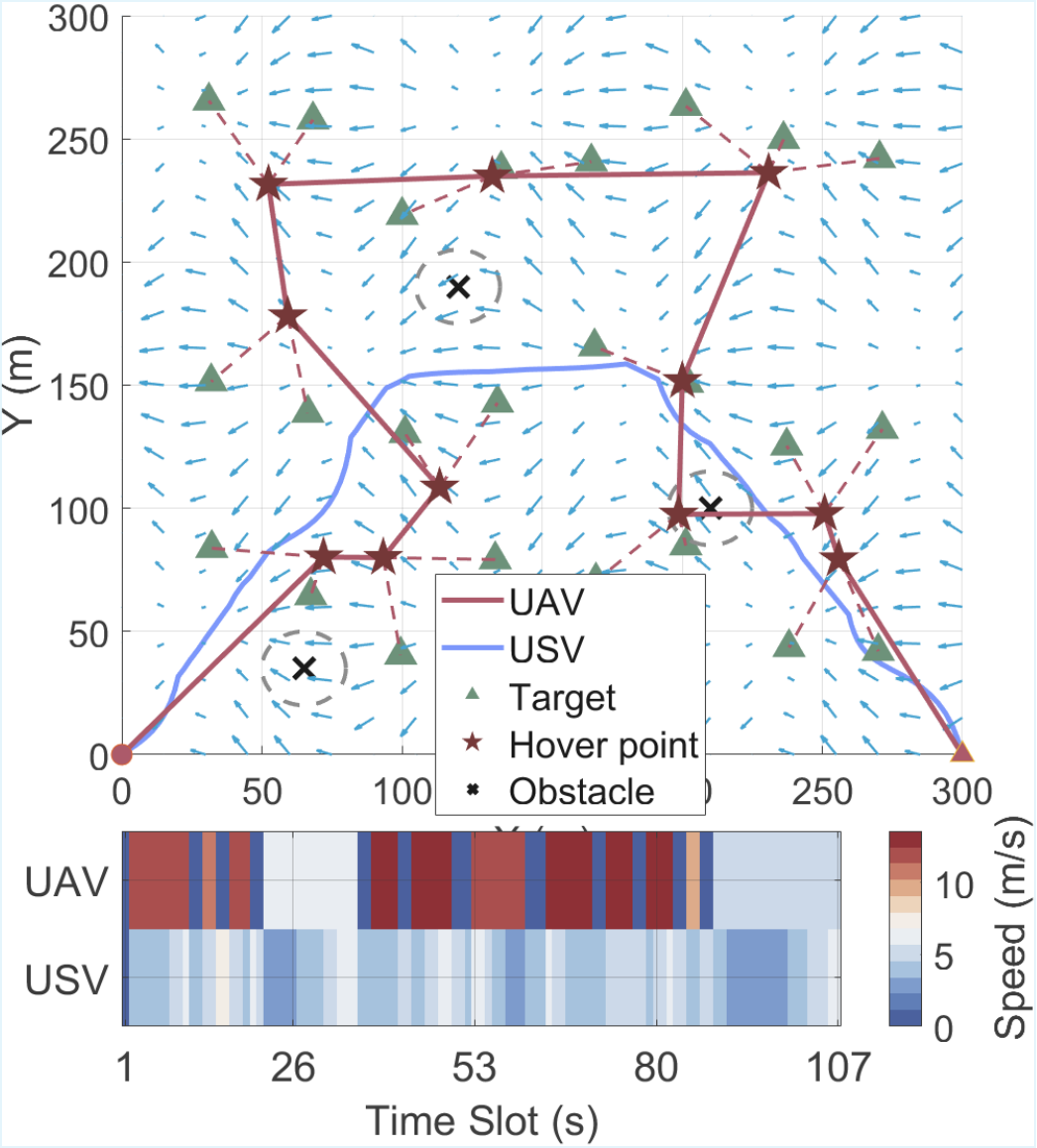}}			
			\caption{The system trajectory at different values of $v_{\text{max}}^w$.}
			\label{WeV} 
		\end{figure}
		
\begin{figure}[!t]
			\centering
			\includegraphics[width=8.8cm]{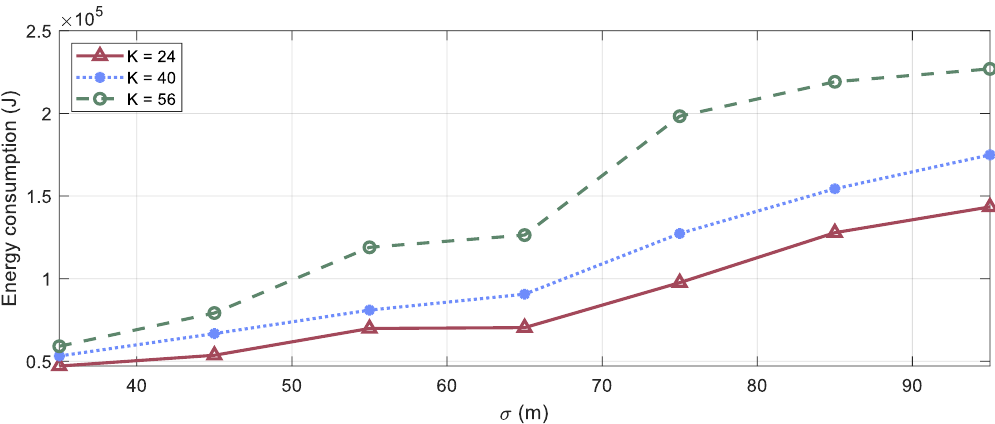}
			\caption{The relationship between energy consumption and $\sigma$.}
			\label{WeKfangcha}
		\end{figure}

		{In Fig.~\ref{WeK_sum}, we provide the trajectory comparisons with different numbers of targets. It can be observed that the trajectory changes substantially with the increasing number of targets. Furthermore, the total duration increases accordingly with $106$s for $K=32$ and $130$s for $K=56$. This result is consistent with the intuition that the UAV and USV require more energy and time duration to guarantee the coverage of the sensing targets.    
			Next, we evaluate the impact of the spatial dispersion of the targets' positions on the trajectory. To this end, the standard deviation $\sigma$ is introduced as  
			\begin{equation}
				\sigma = \sqrt{ \frac{1}{K-1} \sum_{k=1}^{K} (\mathbf{t}_k -\bar{ \mathbf{t} })^2  },
			\end{equation}
			where $\bar{ \mathbf{t}} = \frac{1}{K}\sum_{k=1}^{K}\mathbf{t}_k$ represents the mean of the positions. We can observe in Fig.~\ref{WeK_fangcha} that the degree of deviation can also significantly change the time duration and trajectories of the UAV and USV. Additionally, Fig.~\ref{WeKfangcha} shows that a larger standard deviation leads to a higher energy consumption for the same number of targets. This is because the clustered targets result in compact hover point planning and steady growth in energy consumption. However, the large deviation requires a long path of the UAV and USV's motion, which increases energy consumption.  }

		\subsection{Impact of S\&C Performance Requirements}

\begin{figure}[!t]
			\centering
			\includegraphics[width=9cm]{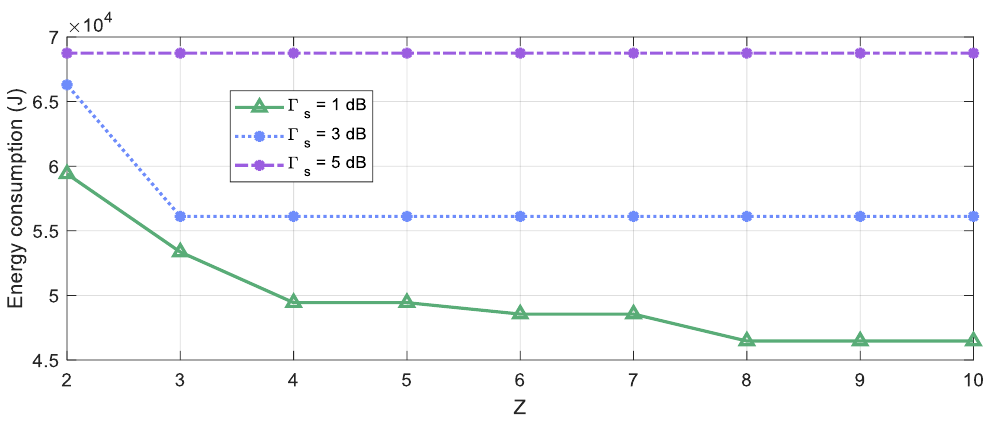}
			\caption{Energy consumption versus maximum number of simultaneously sensed targets $Z$.}
			\label{WeKfangcha1}
		\end{figure}

		{For the sensing process, we first evaluate the relationship between the trajectories and the instantaneous SNR threshold $\Gamma_s$. In Fig.~\ref{WeS}, it is shown that a higher threshold $\Gamma_s$ necessitates closer hover points to each sensing target, thereby increasing the number of hover points. On the other hand, the hover time becomes shorter at each point with $\Gamma_s =5$ dB since the UAV serves fewer targets and hence it is easier to satisfy the total SNR requirement $\Gamma_s^{\text{total}}$. Next, we assess the relationship between the energy consumption and the maximum number of simultaneously sensed targets $Z$ under different values of \( \Gamma _{s} \) in Fig.~\ref{WeKfangcha1}. For a lower SNR threshold (e.g. $\Gamma_s = 1$ dB), energy consumption decreases with an increase in the number of targets detected simultaneously. The main reason is that, on the premise of meeting sensing requirements, an increase in the number of targets served per hover point leads to a reduction in the total number of hover points, consequently decreasing the energy consumption from the kinematic motion of UAV and USV. However, with a higher SNR threshold (e.g. $\Gamma_s = 5$ dB), we find that the energy consumption does not exhibit significant dependency on the number of simultaneously sensed targets (the curve tends to flatten). This phenomenon occurs because the high threshold forces the UAV to approach the targets more closely, preventing it from serving multiple targets simultaneously.        
		}

		\begin{figure}[!t]
			\centering
			\includegraphics[width=9cm]{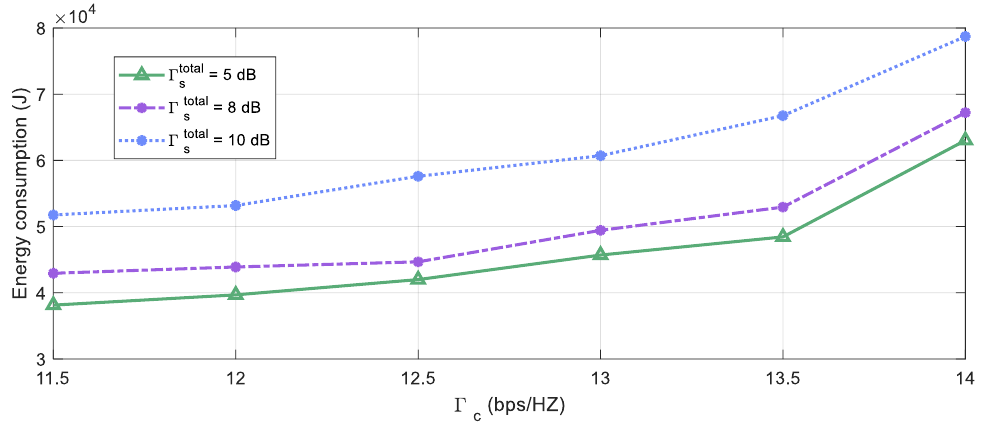}
			\caption{The relationship between energy consumption and $\Gamma_c$.}
			\label{tongxinbianhua}
		\end{figure}
		
		
Fig.~\ref{tongxinbianhua} shows the system energy consumption under different communication rate requirements. As the rate increases, total energy consumption rises significantly due to the UAV adjusting its position and duration to get closer to the USV, while the USV alters its trajectory to approach the UAV. These adjustments meet the communication rate but increase energy usage. Additionally, as the cumulative SNR increases, energy consumption grows because the UAV must extend its sensing hovering time and adjust its trajectory to achieve a higher SNR.

		

		\subsection{Impact of Water Flow and Obstacles}

		{Recalling that the water flow velocity is dependent on the position. In the simulations, we adopt a general complex wave-like water current in \cite{1315730}, where the horizontal and vertical components are
			\begin{equation} \notag
				\begin{aligned}
					v_{x,\mathbf{b}[n]} &= 
					v_{\text{max}}^w(0.8-0.03\sin(0.06b_x[n])\cos(0.03b_y[n])), \\
					v_{y,\mathbf{b}[n]} &=-v_{\text{max}}^w\cos(0.06b_x[n])\cos(0.03b_y[n]),
				\end{aligned}
			\end{equation}
			where \( v_{\text{max}}^w \) represents the maximum water flow speed. Fig. \ref{WeV} illustrates the impact of the water flow on the trajectory. It clearly shows that the water flow can significantly change the trajectory planning, including that of the UAV. This phenomenon will not occur in the Leader-Follower scheme since the UAV trajectory, when designed independently in the leader stage, is unaffected by water currents. In contrast, the trajectories of UAV and USV are jointly designed in the proposed scheme. Furthermore, the energy consumption for Fig. (\ref{w3}) and Fig. (\ref{w-3}) is $62.34$ kJ and $87.20$ kJ, respectively, clearly showing that the energy consumption for upstream navigation is higher than that for downstream navigation. 
		\section{CONCLUSION}
		\label{Sec:6}
		In this paper, we propose an ISAC-empowered UAV-USV collaborative framework for the maritime inspection. The framework jointly optimizes UAV hover points, UAV and USV trajectories, and S\&C beamforming.
		The effectiveness of our framework is validated through experiments, leading to the following insights:
		\begin{itemize}[]
			\item 	
			Under the same number of targets, a more dispersed target distribution (larger standard deviation) leads to a higher rate of energy consumption growth.
			\item
			A higher cumulative SNR notably extends task duration, highlighting the trade-off between performance and efficiency.
			\item 
			A higher communication rate requires the UAV and USV to adjust their speeds and trajectories together, bringing them closer within the same time slot.
		\end{itemize}
		In future research, we will focus on edge computing within this framework and the UAV’s landing and charging process on the USV.
		\appendices
		\section{  }
		\label{appA}
We define the index set \( V = \{1, 2, \ldots, E+2\} \), which includes the virtual base station, start, and end nodes. Let \( s \in V \) and \( t \in V \) denote the start and end node indices, respectively. \( E^\text{cost}_{i,j} \) represents the total cost between the \( i \)-th and \( j \)-th virtual base stations, considering water flow, path length, and velocity, as described in (\ref{cost}).
		Define a binary variable $x_{i,j}$, which satisfies the following
		\begin{equation}
			\begin{aligned}
				x_{i,j}=&
				\begin{cases}
					1, & \text{If a path exists from virtual base} \\
					& \text{station $i$ to virtual base station $j$,} \\
					0, & \text{Otherwise}.
				\end{cases}
			\end{aligned}
		\end{equation}
		
		The shortest path model is defined by
		\begin{subequations}
			\begin{align}
				&\min_{\{ x_{i,j},u_i\}}\sum_{i=0}^{E+2}\sum_{j=0}^{E+2}E^\text{cost}_{i,j} x_{i,j},\: i\neq j, \notag\\
				\mathrm{s.t.}\quad&	\sum^{E+2}_{j=0,j\neq s}x_{s,j}=1,\:\sum^{E+2}_{i=0,i\neq s}x_{i,s}=0,
				\label{tsp1}\\
				&	\sum^{E+2}_{j=0,j\neq t}x_{t,j}=0,\:	\sum^{E+2}_{i=0,i\neq t}x_{i,t}=1,
				\label{tsp2}\\
				\sum^{E+2}_{i=0,i\neq k}&x_{i,k}=1,\:\sum^{E+2}_{j=0,j\neq k}x_{k,j}=1,\: \forall k\neq s,t,
				\label{tsp3}\\
				&	u_{s}=0,\:	1\leq u_i\leq {(E+2)}-1\quad\forall i\neq s,
				\label{tsp4}\\
				u_i-u_j+&\left({E+1}\right)x_{i,j}\leq E,\:\forall i\neq t,j\neq s,i\neq j.
				\label{tsp5}
			\end{align}
		\end{subequations}
		Constraint (\ref{tsp1}) ensures the path starts at node $s$, and Constraint (\ref{tsp2}) ensures it ends at node $t$. 
		Constraint (\ref{tsp3}) ensures each node, except the start and end, is visited once with one entry and exit. 
		Constraints (\ref{tsp4}) and (\ref{tsp5}) eliminate subtours based on the Miller-Tucker-Zemlin (MTZ) formulation, ensuring path connectivity without cycles smaller than $E+2$ \cite{8255824}. 
		The above problem is a Mixed-Integer Linear Programming (MILP) model, solved using Gurobi in MATLAB. 

		\bibliographystyle{IEEEtran}
		\bibliography{reference}

\begin{thebibliography}{10}
\providecommand{\url}[1]{#1}
\csname url@samestyle\endcsname
\providecommand{\newblock}{\relax}
\providecommand{\bibinfo}[2]{#2}
\providecommand{\BIBentrySTDinterwordspacing}{\spaceskip=0pt\relax}
\providecommand{\BIBentryALTinterwordstretchfactor}{4}
\providecommand{\BIBentryALTinterwordspacing}{\spaceskip=\fontdimen2\font plus
\BIBentryALTinterwordstretchfactor\fontdimen3\font minus
  \fontdimen4\font\relax}
\providecommand{\BIBforeignlanguage}[2]{{%
\expandafter\ifx\csname l@#1\endcsname\relax
\typeout{** WARNING: IEEEtran.bst: No hyphenation pattern has been}%
\typeout{** loaded for the language `#1'. Using the pattern for}%
\typeout{** the default language instead.}%
\else
\language=\csname l@#1\endcsname
\fi
#2}}
\providecommand{\BIBdecl}{\relax}
\BIBdecl

\bibitem{11072035}
X.~Ye, Y.~Mao, X.~Yu, S.~Sun, L.~Fu, and J.~Xu, ``Integrated sensing and
  communications for low-altitude economy: A deep reinforcement learning
  approach,'' \emph{IEEE Transactions on Wireless Communications}, pp. 1--1,
  2025.

\bibitem{10879807}
G.~Cheng, X.~Song, Z.~Lyu, and J.~Xu, ``Networked {ISAC} for low-altitude
  economy: Coordinated transmit beamforming and {UAV} trajectory design,''
  \emph{IEEE Transactions on Communications}, pp. 1--1, 2025.

\bibitem{10815625}
Z.~Zhang, L.~Huang, Q.~Wang, L.~Jiang, Y.~Qi, S.~Wang, T.~Shen, B.-H. Tang, and
  Y.~Gu, ``{UAV} hyperspectral remote sensing image classification: A
  systematic review,'' \emph{IEEE Journal of Selected Topics in Applied Earth
  Observations and Remote Sensing}, vol.~18, pp. 3099--3124, 2025.

\bibitem{10418158}
B.~He, X.~Ji, G.~Li, and B.~Cheng, ``Key technologies and applications of
  {UAV}s in underground space: A review,'' \emph{IEEE Transactions on Cognitive
  Communications and Networking}, vol.~10, no.~3, pp. 1026--1049, 2024.

\bibitem{10872967}
Z.~Zhang, R.~He, B.~Ai, M.~Yang, X.~Zhang, Z.~Qi, and Y.~Yuan, ``Channel
  measurements and modeling for dynamic vehicular {ISAC} scenarios at 28 ghz,''
  \emph{IEEE Transactions on Communications}, pp. 1--1, 2025.

\bibitem{10845869}
F.~Dong, F.~Liu, S.~Lu, Y.~Xiong, Q.~Zhang, Z.~Feng, and F.~Gao,
  ``Communication-assisted sensing in 6g networks,'' \emph{IEEE Journal on
  Selected Areas in Communications}, vol.~43, no.~4, pp. 1371--1386, 2025.

\bibitem{10566041}
Y.~Liu, X.~Liu, Z.~Liu, Y.~Yu, M.~Jia, Z.~Na, and T.~S. Durrani, ``Secure rate
  maximization for {ISAC}-{UAV} assisted communication amidst multiple
  eavesdroppers,'' \emph{IEEE Transactions on Vehicular Technology}, vol.~73,
  no.~10, pp. 15\,843--15\,847, 2024.

\bibitem{10752639}
A.~Li, G.~Guan, H.~Zhao, S.~Li, J.~Zhu, X.~Han, Y.~Wang, and J.~Pan,
  ``Integrated methodology for atmospheric correction and cloud removal of
  multispectral remote sensing images using guided diffusion model,''
  \emph{IEEE Transactions on Geoscience and Remote Sensing}, vol.~62, pp.
  1--21, 2024.

\bibitem{10499863}
Y.~Liao, Y.~Song, S.~Xia, Y.~Han, N.~Xu, X.~Zhai, and Z.~Yuan, ``Low-latency
  data computation of inland waterway {USV}s for ris-assisted {UAV} mec
  network,'' \emph{IEEE Internet of Things Journal}, vol.~11, no.~16, pp.
  26\,713--26\,726, 2024.

\bibitem{10529184}
X.~Jing, F.~Liu, C.~Masouros, and Y.~Zeng, ``{ISAC} from the sky: {UAV}
  trajectory design for joint communication and target localization,''
  \emph{IEEE Transactions on Wireless Communications}, vol.~23, no.~10, pp.
  12\,857--12\,872, 2024.

\bibitem{10295964}
X.~Liu, Y.~Liu, Z.~Liu, and T.~S. Durrani, ``Fair integrated sensing and
  communication for multi-{UAV}-enabled internet of things: Joint 3-d
  trajectory and resource optimization,'' \emph{IEEE Internet of Things
  Journal}, vol.~11, no.~18, pp. 29\,546--29\,556, 2024.

\bibitem{10100680}
C.~Deng, X.~Fang, and X.~Wang, ``Beamforming design and trajectory optimization
  for {UAV}-empowered adaptable integrated sensing and communication,''
  \emph{IEEE Transactions on Wireless Communications}, vol.~22, no.~11, pp.
  8512--8526, 2023.

\bibitem{9847217}
S.~Hu, X.~Yuan, W.~Ni, and X.~Wang, ``Trajectory planning of cellular-connected
  {UAV} for communication-assisted radar sensing,'' \emph{IEEE Transactions on
  Communications}, vol.~70, no.~9, pp. 6385--6396, 2022.

\bibitem{10713326}
S.~Peng, B.~Li, L.~Liu, Z.~Fei, and D.~Niyato, ``Trajectory design and resource
  allocation for multi-{UAV}-assisted sensing, communication, and edge
  computing integration,'' \emph{IEEE Transactions on Communications}, vol.~73,
  no.~4, pp. 2847--2861, 2025.

\bibitem{10680299}
A.~Khalili, A.~Rezaei, D.~Xu, F.~Dressler, and R.~Schober, ``Efficient {UAV}
  hovering, resource allocation, and trajectory design for {ISAC} with limited
  backhaul capacity,'' \emph{IEEE Transactions on Wireless Communications},
  vol.~23, no.~11, pp. 17\,635--17\,650, 2024.

\bibitem{10769423}
Y.~Liu, W.~Mao, B.~He, W.~Huangfu, T.~Huang, H.~Zhang, and K.~Long, ``Radar
  probing optimization for joint beamforming and {UAV} trajectory design in
  {UAV}-enabled integrated sensing and communication,'' \emph{IEEE Transactions
  on Communications}, pp. 1--1, 2024.

\bibitem{10787434}
Y.~Wu, Y.~Sun, X.~Yu, D.~Zhang, and W.~He, ``Intelligent experiment robotic
  systems design for material preparation and detection,'' \emph{IEEE
  Transactions on Systems, Man, and Cybernetics: Systems}, vol.~55, no.~2, pp.
  1241--1251, 2025.

\bibitem{9944188}
W.~Wei, J.~Wang, Z.~Fang, J.~Chen, Y.~Ren, and Y.~Dong, ``3u: Joint design of
  {UAV}-{USV}-uuv networks for cooperative target hunting,'' \emph{IEEE
  Transactions on Vehicular Technology}, vol.~72, no.~3, pp. 4085--4090, 2023.

\bibitem{10643681}
J.~Yan, J.~Lin, X.~Yang, C.~Chen, and X.~Guan, ``Cooperation detection and
  tracking of underwater target via aerial–surface–underwater vehicles,''
  \emph{IEEE Transactions on Automatic Control}, vol.~70, no.~2, pp.
  1068--1083, 2025.

\bibitem{10530448}
N.~Wang, X.~Liang, Z.~Li, Y.~Hou, and A.~Yang, ``Pse-d model-based cooperative
  path planning for {UAV} and {USV} systems in antisubmarine search missions,''
  \emph{IEEE Transactions on Aerospace and Electronic Systems}, vol.~60, no.~5,
  pp. 6224--6240, 2024.

\bibitem{9085942}
Y.~Wu, K.~H. Low, and C.~Lv, ``Cooperative path planning for heterogeneous
  unmanned vehicles in a search-and-track mission aiming at an underwater
  target,'' \emph{IEEE Transactions on Vehicular Technology}, vol.~69, no.~6,
  pp. 6782--6787, 2020.

\bibitem{10423261}
X.~Cao, W.~Liu, and L.~Ren, ``Underwater target capture based on heterogeneous
  unmanned system collaboration,'' \emph{IEEE Transactions on Intelligent
  Vehicles}, vol.~9, no.~10, pp. 6049--6062, 2024.

\bibitem{9453748}
W.~Wang, X.~Li, R.~Wang, K.~Cumanan, W.~Feng, Z.~Ding, and O.~A. Dobre,
  ``Robust 3d-trajectory and time switching optimization for dual-{UAV}-enabled
  secure communications,'' \emph{IEEE Journal on Selected Areas in
  Communications}, vol.~39, no.~11, pp. 3334--3347, 2021.

\bibitem{liu2023fair}
X.~Liu, Y.~Liu, Z.~Liu, and T.~S. Durrani, ``Fair integrated sensing and
  communication for multi-{UAV}-enabled internet of things: Joint 3-d
  trajectory and resource optimization,'' \emph{IEEE Internet of Things
  Journal}, vol.~11, no.~18, pp. 29\,546--29\,556, 2023.

\bibitem{9124713}
X.~Liu, T.~Huang, N.~Shlezinger, Y.~Liu, J.~Zhou, and Y.~C. Eldar, ``Joint
  transmit beamforming for multiuser mimo communications and mimo radar,''
  \emph{IEEE Transactions on Signal Processing}, vol.~68, pp. 3929--3944, 2020.

\bibitem{1315730}
A.~Alvarez, A.~Caiti, and R.~Onken, ``Evolutionary path planning for autonomous
  underwater vehicles in a variable ocean,'' \emph{IEEE Journal of Oceanic
  Engineering}, vol.~29, no.~2, pp. 418--429, 2004.

\bibitem{8255824}
Y.~Zeng, X.~Xu, and R.~Zhang, ``Trajectory design for completion time
  minimization in {UAV}-enabled multicasting,'' \emph{IEEE Transactions on
  Wireless Communications}, vol.~17, no.~4, pp. 2233--2246, 2018.

\bibitem{5447068}
Z.-q. Luo, W.-k. Ma, A.~M.-c. So, Y.~Ye, and S.~Zhang, ``Semidefinite
  relaxation of quadratic optimization problems,'' \emph{IEEE Signal Processing
  Magazine}, vol.~27, no.~3, pp. 20--34, 2010.

\end{thebibliography}

	\end{document}